\documentclass[useAMS,usenatbib]{mn2e}
\usepackage{graphicx}
\usepackage{txfonts}
\usepackage{wrapfig}
\usepackage{longtable}
\usepackage[usenames]{color}

\def\aj{AJ}%
\def\araa{ARA\&A}%
\def\apj{ApJ}%
\def\apjs{ApJS}%
\def\aap{A\&A}%
\def\aapr{A\&A~Rev.}%
\def\mnras{MNRAS}%
\def\prd{Phys.~Rev.~D}%
\def\pasp{PASP}%
\newcommand{\bc}{\begin{center}}
\newcommand{\ec}{\end{center}}

\begin{document}
\title[Evidence for Inside-out Formation of Galactic Disks]
      {The GALEX Arecibo SDSS survey: III. Evidence for the Inside-Out Formation of Galactic Disks}
\author[Jing ~Wang et al.]
       {Jing Wang$^{1,2}$\thanks{Email: wangj@mpa-garching.mpg.de},
       Guinevere Kauffmann$^2$, Roderik Overzier$^2$, Barbara Catinella$^2$,
       \newauthor David Schminovich$^3$, Timothy M. Heckman$^4$, Sean M. Moran$^4$, Martha P. Haynes$^5$ 
       \newauthor Riccardo Giovanelli$^5$, Xu Kong$^{1,6}$ \\
       $^1$Center for Astrophysics, University of Science and Technology of China,
        230026 Hefei, China
       \\
       $^2$Max--Planck--Institut f\"ur Astrophysik,
        Karl--Schwarzschild--Str. 1, D-85748 Garching, Germany
        \\
       $^3$Department of Astronomy, Columbia University, New York, NY 10027, USA\\
       $^4$Department of Physics and Astronomy, The Johns Hopkins University, Baltimore, MD 21218, USA\\
       $^5$Center for Radiophysics and Space Research, Cornell University, Ithaca, NY 14853, USA\\
       $^6$Key Laboratory for Research in Galaxies and Cosmology, 
   University of Science and Technology of China, Chinese Academy of 
   Sciences, China}

\date{Accepted by MNRAS;
      in original form 2010 September 01}

\pubyear{2010}

\maketitle

\begin{abstract}
We analyze a sample
of galaxies with stellar masses greater than $10^{10} M_{\odot}$ and with redshifts
in the range $0.025<z<0.05$  for which HI
mass measurements are available  from the GALEX Arecibo SDSS Survey (GASS) or from the Arecibo Legacy Fast ALFA survey
(ALFALFA). At a given value of $M_*$, our sample consists primarily of
galaxies that are more HI-rich than average.                         
We constructed a series of three control samples for comparison with these HI-rich galaxies:
one sample is  matched in  stellar mass
and redshift (C$_{M*}$),  the second sample is matched in  stellar
mass, $NUV-r$ colour and redshift (C$_{M*,NUV-r}$), and the third sample
is matched in stellar mass, $NUV-r$ colour, stellar surface mass 
density $\mu_*$ and redshift.  We generated
self-consistent 7-band photometry  (FUV, NUV, $u,g,r,i,z$)
for all galaxies  and we used this to derive inner
colours, outer colours, asymmetry and smoothness parameters.
We also used standard SED fitting techniques to  derive  inner and outer specific star formation rates.
As  expected, HI-rich
galaxies differ strongly from galaxies of 
same stellar mass that are selected without regard
to HI content. The majority  of these
differences are  attributable to the fact that galaxies with more
gas are bluer and more actively star-forming. 
In order to identify those  galaxy properties that are {\em  causally connected
with HI content}, we  compare
results derived for the HI sample with those derived for galaxies 
matched in stellar mass, size and NUV-$r$ colour. 
The only photometric property  that is clearly
attributable to increasing HI content,
is the  colour gradient of the galaxy.
Galaxies with larger HI fractions have bluer, more actively star-forming outer disks   
compared  to the inner part of the galaxy. HI-rich galaxies also have larger  
$g$-band radii compared to $i$-band radii.   
Our results are consistent with the  ``inside-out'' picture of disk galaxy formation, which has commonly served
as a basis for semi-analytic models of  the formation of disks
in the context of  Cold Dark Matter cosmologies. The lack of any intrinsic connection between  HI fraction and  galaxy 
asymmetry suggests that gas is accreted smoothly onto the outer disk.

\end{abstract}

\section{Introduction}
The fraction of the baryons in dark matter halos that are locked up in stars in the
central galaxy reaches a maximum of 20\%, at masses
somewhat below that of the Milky Way, and falls rapidly at both higher and
lower masses (Guo \& White 2010). In the absence of feedback effects from supernova or 
AGN, most of these baryons are expected to   
cool, accrete and form stars \citep{Kauffmann93}. 

There is some observational evidence that  gas from the
external environment continues to accrete onto galaxies at the present day.
Without a continuous supply of gas, most
star forming galaxies would run out of gas on timescales of a few Gyr
\citep{Larson80}.  
HI cloud complexes, HI-rich dwarfs in the vicinity of spiral galaxies,
extended and warped outer layers of HI in spiral galaxies, and lopsided
galaxy disks have all been cited as evidence for ongoing gas accretion in nearby galaxies
\citep{Sancisi08}.  However, estimates of the accretion rates in the form of
neutral hydrogen in nearby spirals consistently give values much lower than those 
required to sustain star formation at their observed rates (see Fraternali
(2010) for a recent review).  This leads to the hypothesis that most  of the
baryons that reside outside galaxies at the present  are in the form of ionized gas (e.g. Fukugita, Hogan \&
Peebles 1998).  There is also some evidence from observations of  ionized silicon in high
and intermediate-velocity clouds in the Milky Way, that gas in a
low-metallicity ionized phase in the halo can provide a substantial (1
M$_{\odot}$/yr) cooling inflow to replenish star formation in the disk
(Shull et al. 2009).

One interesting question that has not been considered very much in the
literature, is whether gas accretes onto all galaxies in a smooth,
continuous fashion, or whether accretion is more stochastic. If accretion
has been stochastic, then galaxies that have recently acquired gas from the
external environment might be expected to have  unusually high HI mass
fractions.
The neutral gas content of a galaxy  has long been  known to correlate with
its other physical properties. HI mass fractions increase smoothly along
the Hubble sequence from the early-type (S0) to the late-type (Im) end
\citep{RH94}.  There are also  correlations between HI mass fraction and
galaxy properties such as stellar mass ($M_*$), stellar mass surface density
($\mu_*$)  and
broadband colour.  These correlations  form the basis for the so-called
``photometric gas fraction'' technique for predicting the HI  gas content 
galaxies \citep{Kannappan04, Zhang09}.
Rather little attention has been given  to the {\em scatter} in gas
fraction at a given value of $M_*$, $\mu_*$ or colour.    
Recently, Zhang et al (2009) found that more gas-rich galaxies were systematically deficient in metals 
at fixed stellar mass and they interpreted this as evidence for recent
cosmological infall of gas in these systems.  However, up to now  there has been
no systematic study of how galaxy properties vary as a function of 
HI over- or under-abundance.  

One  problem that has hampered progress on this front 
has been the lack of suitable data sets. Blind HI
surveys, such as the HI Parkes All Sky Survey (HIPASS, Barnes et al 2001)
or the Arecibo Legacy Fast ALFA survey (ALFALFA, Giovanelli et al 2005)
have produced large,  unbiased samples of galaxies selected by
HI mass. However, these surveys are still relatively  shallow, so the
fraction of HI-poor galaxies is small. 

Recently, Catinella et al (2010, hereafter C10) explored how
HI gas fraction scales as a function of galaxy stellar mass,
galaxy structural parameters and NUV-r colour, using data from the GALEX Arecibo SDSS
Survey (GASS).  The survey is an ongoing large programme at the Arecibo
radio telescope that is gathering high quality HI-line spectra for an
unbiased sample of $\sim 1000$ galaxies with stellar masses greater than
$10^{10} M_{\odot}$
 and redshifts in the range  0.025$<z<$ 0.05, selected from the
Sloan Digital Sky Survey (SDSS, \citet{York00}) spectroscopic and Galaxy Evolution Explorer
(GALEX, \citet{Martin05}) imaging surveys. The galaxies are observed until detected or until
a low gas mass fraction limit (1.5-5 \% ) is reached.  C10 quantify in
detail how the {\em mean} atomic gas fraction of the galaxies in their
sample  decreases as a function of stellar mass, stellar mass surface
density, galaxy bulge-to-disc ratio (as measured by the concentration index
of the r-band light), and global  NUV-r colour. The fraction of galaxies
with significant (more than a few percent) HI decreases sharply above a
characteristic stellar surface mass density of $10^{8.5}$ M$_{\odot}$
kpc$^{-2}$.  The fraction of gas-rich galaxies decreases much more smoothly
with stellar mass.

In this paper, we extend the study of C10 to study how the {\em resolved}
UV/optical photometric  properties of galaxies depend on atomic gas
fraction.  We have developed a photometric pipeline that transforms the
SDSS and GALEX images to the same geometry and effective resolution, so that
our photometric measurements trace the same part of each galaxy at
different wavelengths (Wang et al 2009). We use these 
images to study UV-optical colours in
their inner and outer regions of the galaxies in our sample and to study
how colour (and hence specific star formation rate) {\em gradients} depend
on HI content.  
In order to 
extend the range in M(HI)/$M_*$ that we are able to probe, we
combine the sample analyzed in C10 with 
galaxies  from the GASS ``parent sample''  (described in
Section~\ref{sec:data}) which have  HI detections in the ALFALFA survey.
This considerably  improves the statistics for galaxies with large values of  M(HI)/$M_*$,
which are relatively rare. In order to isolate the effect of the presence or absence
of gas on the photometric properties of galaxies,  
 we compare our results with three control samples,
which have similar optical properties to the galaxies in our sample
with HI mass measurements, but are selected without 
regard to HI content. These control samples   
are described in detail in Section~\ref{subsec:controlsample}.  Sections
~\ref{subsec:phot_tech} and ~\ref{subsec:SF_tech} 
present the methods used  to measure our
photometric parameters, and in Section 4 we discuss the trends in galaxy size, $NUV-r$
and specific star formation rate gradients, and higher order 
morphological parameters such as asymmetry and smoothness as a 
function of  HI mass fraction.
Finally Section~\ref{sec:summary} summarizes  our results. Throughout this paper,
we assume a cosmology with $H_0$=70 km s$^{-1}$ Mpc$^{-1}$, $\Omega_m$=0.3,
and  $\Omega_\Lambda$=0.7 \citep{Tegmark04}.  A Chabrier initial mass
function is assumed in the stellar population synthesis analysis
\citep{Chabrier03}.

\section{Data}
\label{sec:data}
\subsection{The parent sample}
\label{subsec:PSsample}
We select galaxies with stellar masses $M_*>10^{10}M_{\odot}$ in the redshift range
$0.025<z<0.05$  from the sixth data release (DR6) of the SDSS Survey,
which lie within the maximal ALFALFA footprint. 
We match this catalogue with the fourth data release (GR4) of the GALEX survey
and this yields a sample of 10468 galaxies. We further limit the sample to face-on galaxies with 
$b/a>0.4$ (or an inclination of less than 66.4 deg) where $b$ and $a$ are the minor and major axes of an ellipsoidal fit (from SDSS $r$ band) to each galaxy.
This yields a final  sample of 8429 galaxies, which we refer to as the {\em parent sample} from
now on. We restrict the sample to face-on systems,
so that our estimates of
star formation rate, which are derived from the UV/optical photometry,
will be less affected by  dust attenuation effects (Section~\ref{subsec:SF_tech}). 

\subsection { The sample of galaxies with HI mass measurements}
\label{subsec:HIsample}
\begin{figure*}
 \centering
  \includegraphics[width=6cm]{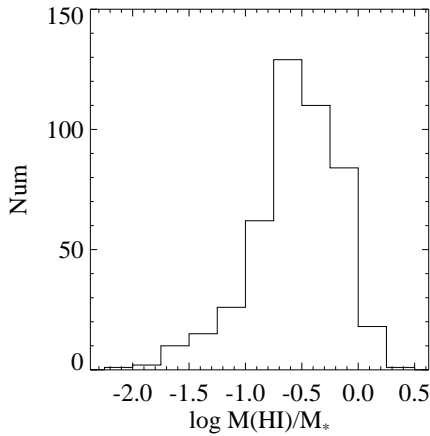}
 \caption{The distribution of HI mass fractions for the 458 galaxies in our sample with  HI mass measurements. }
 \label{fig:HI_frac}
\end{figure*}

\begin{figure*}
 \centering
 \includegraphics[width=17cm]{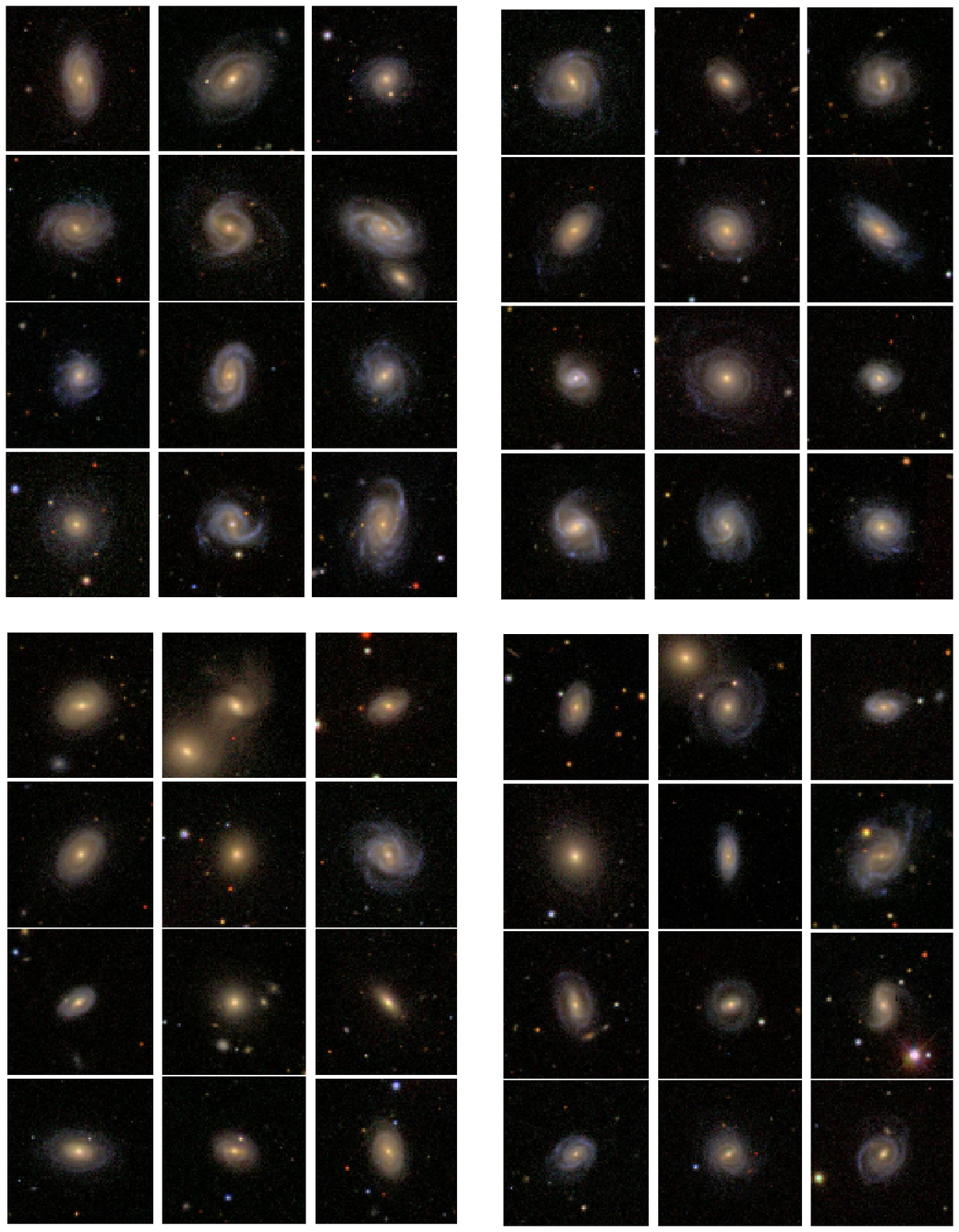}
 \caption{A montage of $g,r,i$ colour images of  galaxies selected
from the HI sample (top-left panel), and the corresponding control galaxies from the
C$_{M*}$ (bottom-left panel), C$_{M*,NUV-r}$ (top-left panel) and  C$_{M*,NUV-r,\mu_*}$ (bottom-right panel) samples.
The scale of each image is $\sim$0.2$''$ pixel$^{-1}$. The galaxies are selected to
have stellar masses in the range from  10$^{10.75}$ to 10$^{11.5}$ M$_{\odot}$.
The galaxies from the HI sample are ordered by increasing  $\log M(HI)/M_*$, from -0.84 to -0.48}
 \label{fig:gallery}
\end{figure*}

This sample consists of galaxies in the parent sample for which we have catalogued HI detections from the $40\%$ ALFALFA survey ($\alpha.40$, which will be discussed in \citet{Martin10} and \citet{Haynes10}), as well as galaxies
that are included in the  first data release (DR1) of the GASS survey (see C10 for a detailed description). Our HI sample consists of 
458 galaxies with a median HI mass fraction 
M(HI)/M$_*$ of $\sim$ 30$\%$. Figure~\ref{fig:HI_frac} shows the distribution of M(HI)/M$_*$ for
these galaxies. The left-top panel of Figure~\ref{fig:gallery} show       
a  gallery of $g,r,i$ colour images generated using  the SDSS Image List Tool \footnote[1]{http://cas.sdss.org/astro/en/tools/chart/list.asp}
for galaxies with stellar masses in the range 
 $10^{10.75}$ to $10^{11.5}$  M$_{\odot}$ and  HI mass fractions 
($\log$ M(HI)/$M_*$) in the range  -0.84 to -0.48 dex.
We note that the 
average value of $\log$ M(HI)/$M_*$ for galaxies
in this stellar mass range is   $\sim$-1.4 (see C10), so 
these galaxies are all significantly more HI-rich than is typical. 

\subsection{Control samples}
\label{subsec:controlsample}
We limit the galaxy pool from which we draw the 
control galaxies  to the regions of sky not yet covered by
our current  ALFALFA or GASS catalogues.
The sample from which we draw the control
galaxies comprises 5527 galaxies from the ``parent sample''. We create three control galaxy samples with similar optical
properties to the galaxies with HI detections.

The first control sample ($C_{M*}$) is matched only in redshift and stellar mass. 
For each galaxy with an HI detection, we search within a distance of 0.028 dex in $M_*$ and 0.005 in $z$ for
the nearest neighbor in the plane of stellar mass versus
redshift. Each galaxy only enters the control sample once. We 
repeat this procedure twice, increasing the 
$\Delta \log M_*$ tolerance to 0.03 and 0.035, so that
in the end, each galaxy with an HI-detection has 3 galaxies 
matched in stellar mass and redshift.  
The bottom-left panel of Figure~\ref{fig:gallery} shows a montage of  $C_{M*}$ galaxies
that have been matched to galaxies in Rows 1-2.
Note that because the control galaxies are matched in redshift,
we can meaningfully compare the morphologies, sizes 
and colours of the HI rich and corresponding $C_{M*}$ galaxies
in this figure.

\begin{figure*}
\centering
\includegraphics[width=12.5cm]{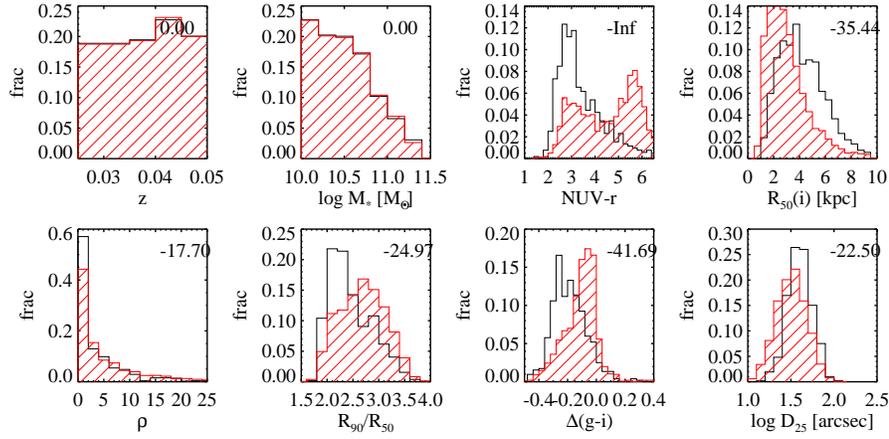}
\caption{Comparison of distributions of redshifts ($z$), stellar masses (log M$_*$), total $NUV-r$ colours, half-light radii ($R_{50}$), local
environment densities ($\rho$), concentration indices ($R_{90}/R_{50}$), $g-i$ colour differences ($\Delta(g-i)$) and 25 mag arcsec$^{-2}$ isophote diameters (D$_{25}$)  for  the sample with HI measurements  (open histograms) and the $C_{M*}$
control sample  (red, hatched histograms). The logarithm of KS-test probabilities that the two
histograms are drawn from the same underlying distribution are denoted in the corner of each panel.}
\label{fig:quality_lgm}
\end{figure*}

Figure~\ref{fig:quality_lgm} compares the distributions 
of redshifts ($z$), stellar masses (log M$_*$), total $NUV-r$ colours, half-light radius ($R_{50}$), local
environment densities \footnote {Note that $\rho$ is calculated as the number of tracer galaxies inside a cylinder of diameter 2 Mpc h$^{-1}$ and
length  16 Mpc h$^{-1}$ divided by the mean number of galaxies inside such a cylinder 
at that redshift.  The $\rho$ values have been extracted  from the NYU Value-Added Galaxy Catalog \citep{Blanton05}} ($\rho$), concentration indices ($R_{90}/R_{50}$), $g-i$ colour difference ($\Delta(g-i)$, see Sect.3.1) and 25 mag arcsec$^{-2}$ isophote diameters (D$_{25}$) for galaxies with  HI measurements (open histograms) and
for  $C_{M*}$ galaxies (hatched, red histograms).
The logarithm of the  KS-test probabilities that both histograms are drawn from the same 
underlying distribution  are noted in  the corner of each panel: a value of -2 means there is 99$\%$ confidence 
that the null hypothesis that the two samples 
are drawn from the same distribution can be rejected.

Figure~\ref{fig:quality_lgm} confirms the visual impression gained
from comparing  the left-top and left-bottom panels of Figures ~\ref{fig:gallery}: 
Galaxies in our sample with measured HI masses are, on average, significantly more 
gas rich than $C_{M*}$ galaxies and are also found to have 
bluer global colours, lower concentrations, larger sizes, and stronger negative
colour differences (i.e. they are bluer on the outside).   
Galaxies with HI measurements
are found in lower density environments, which is not 
too surprising, since galaxy colour and environment
are known to be strongly correlated.

We have  constructed a second control sample, $C_{M*,NUV-r}$, that is matched in $NUV-r$ colour in addition to stellar mass and redshift. 
For each galaxy with an HI measurement, 
we search within a distance of 0.18 dex in $\log $ M$_*$, 0.31 mag in $NUV-r$ colour  and 0.005 in $z$ for 
the nearest neighbor in the 
stellar mass-redshift-colour plane. Note that the chosen 
tolerances in  $\log $ M$_*$
and  $NUV-r$ colour are comparable to the 
errors in these quantities. We repeat the above 
process again, with the log M$_*$ 
distance increased to 0.25 and the ($NUV-r$) distance 
increased to 0.43 mag. In the end each HI galaxy has 
2 control galaxies matched in stellar mass and ($NUV-r$) colour.

\begin{figure*}
\centering
\includegraphics[width=12.5cm]{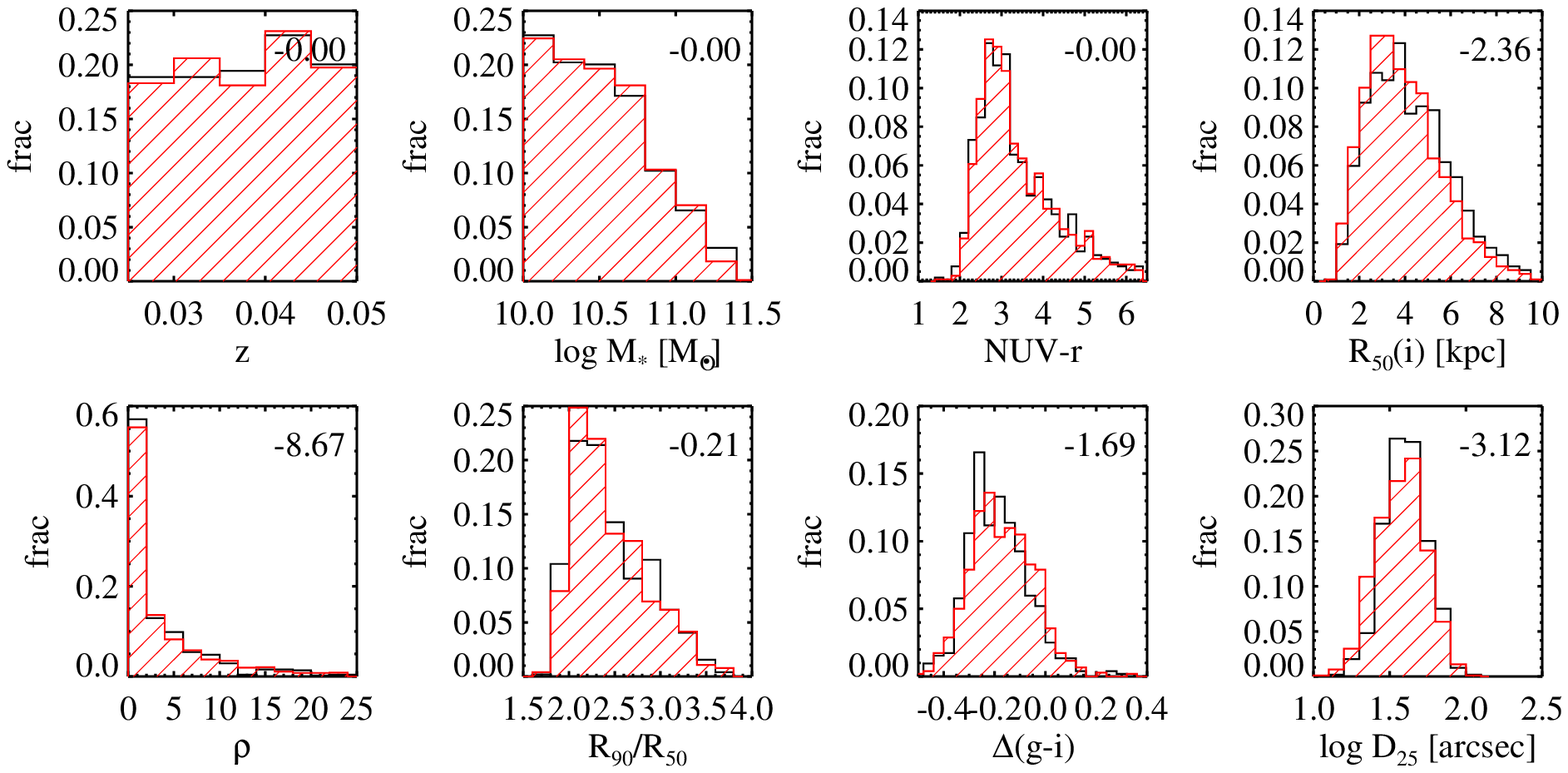}
\caption{Same as Figure~\ref{fig:quality_lgm}, but for the C$_{M*,NUV-r}$ control sample.}
 \label{fig:quality_lgm_nuvr}
\end{figure*}

The  $C_{M*,NUV-r}$ control sample (examples are shown in the 
right-top panel of  Figure~\ref{fig:gallery})  allows us to investigate the extent to which the $NUV-r$ colour
can serve as a ``proxy'' for the HI content of a galaxy. 
Note that this is the underlying assumption of many
photometric gas fraction techniques -- if HI fraction and colour are exactly equivalent, then the HI
sample and the $C_{M*,NUV-r}$ control sample should have identical
properties. 
Figure~\ref{fig:quality_lgm_nuvr} presents 
histograms of the properties of these two samples.
As can be seen, there is no longer any significant difference in concentration
index between the sample with HI masses and the $C_{M*,NUV-r}$ control sample,
but the size differences do persist. The differences in the distribution of local
environment parameters and colour gradients decrease significantly, but
are still significant.  

Finally, we have  constructed a third control sample, $C_{M*,NUV-r,
\mu*}$, that is matched in half-light radius measured from the SDSS $i$-band image ($R_{50}(i)$) in addition to stellar mass, $NUV-r$ colour and redshift. 
For each galaxy with an HI measurement, 
we search within a distance of 0.3 dex in $\log $ M$_*$, 0.52 mag in $NUV-r$ colour, 1.9 kpc in $R_{50}(i)$, and 0.005 in $z$ for 
the nearest neighbor in the 
stellar mass-redshift-colour-size plane.  In this case, there is only
one control galaxy matched to each  HI galaxy. 
Catinella et al (2010) showed that a HI mass could be most accurately ``predicted''
using a linear combination of NUV$-r$ colour and $\log \mu_*$.  
Figure 5 shows that very few apparent differences between the HI sample and the
the  $C_{M*,NUV-r,\mu_*}$ control sample now remain. There is no longer any significant
difference in the distributions of local environment parameters.   
There does appear to be a difference between the distribution of $g-i$ colour
gradients, which we will investigate in more detail in the next section\footnote{ We also made a control sample similar to $C_{M*,NUV-r}$, but which excluded galaxies hosting active or composite nuclei (AGN)
using the criteria in \citet{Kauffman03agn}. We did not find any difference in the results.}.

\begin{figure*}
\centering
\includegraphics[width=12.5cm]{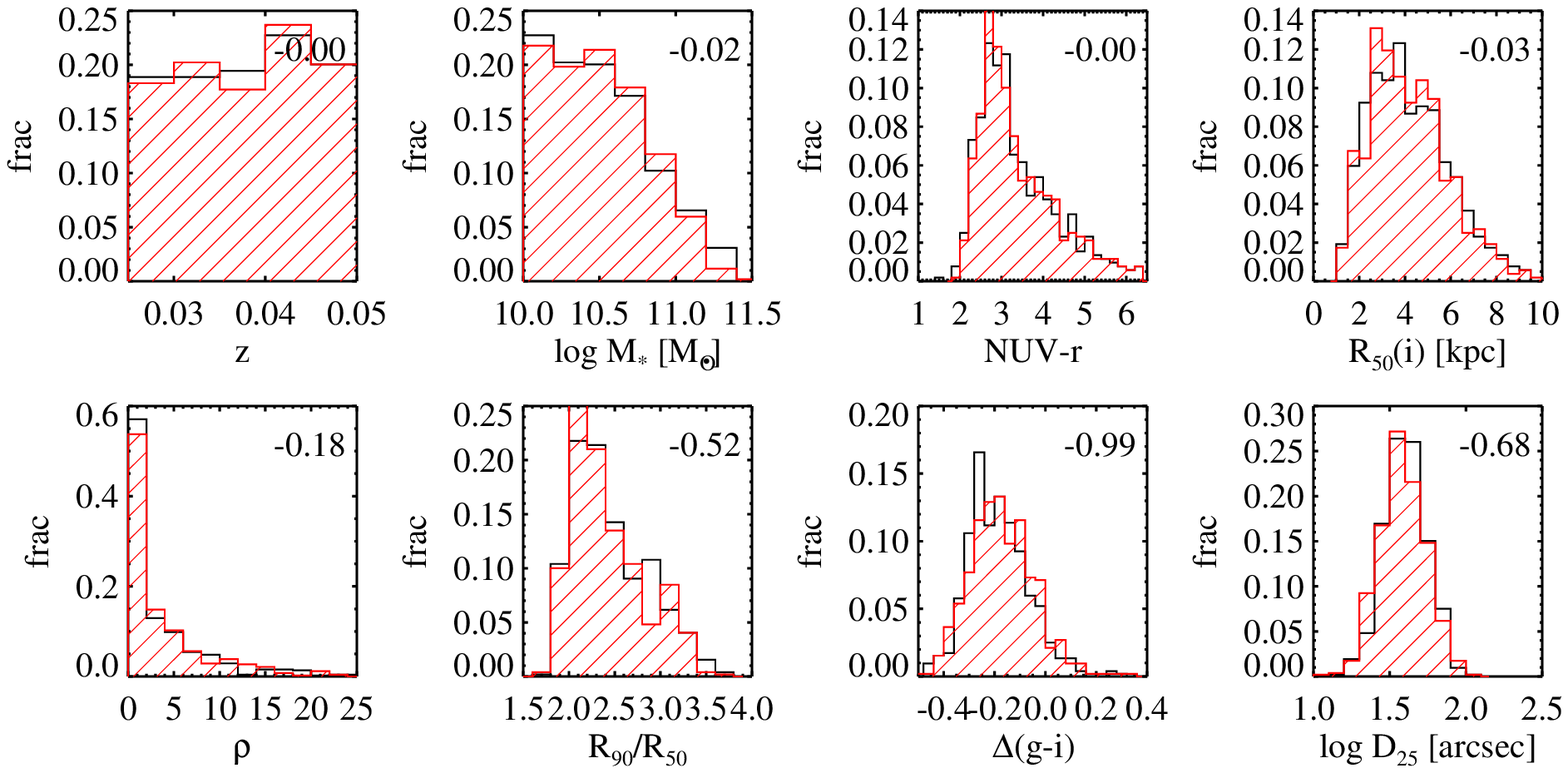}
\caption{Same as Figure~\ref{fig:quality_lgm}, but for the C$_{M*,NUV-r,\mu*}$ control sample.}
 \label{fig:quality_lgm_nuvr_mu}
\end{figure*}

\section{Measurements}
\label{sec:measurements}
\subsection{2-zone photometry}
\label{subsec:phot_tech}
In this section, we analyze how the sizes and colour gradients of galaxies of fixed stellar mass depend on atomic
gas fraction.
We have generated two different sets of  aperture-matched photometry
for galaxies with  HI measurements and also for the galaxies in our control samples.
The first set of measurements is generated from GALEX FUV and NUV images, and SDSS images in the $u,g,r,i$ and $z$-bands.
Images in the  FUV, $u, g, r, i$, and  $z$ bands are registered so that they have  the same geometry as 
the NUV image, and are then  convolved to the same effective resolution as  the NUV image.
All neighboring objects  are masked and the
$r$-band image is used  as the reference for flux measurements in all other bands (see \citet{wang09} for details).
We use this set of photometric measurements in our analysis of UV/optical colours and colour gradients,
and for self-consistent SED fitting using all 7 photometric bands (FUV,NUV, $u,g,r,i,z$).  

The second set of measurements is generated from the SDSS $g$ and $i$ band images, which
have significantly higher resolution than the GALEX images ($\sim 1$ $''$ instead of 5 $''$). 
The $i$-band image is registered to
match the geometry of the $g$ band image, and the two images are convolved so that they match to a common PSF.
The $g$-band image is used as the reference image for flux measurements in both bands. 
This set of photometric measurements is used whenever we consider only optical colours, 
sizes or colour gradients.  

We divide each galaxy into an inner and an outer region: the former is enclosed by $R_{50}$,  the 
radius enclosing half the total $r$-band light  (determined from the PSF-convolved $r$ band image),
and the latter is defined as the region between $R_{50}$  and 2.5 times the Kron radius 
(\citet{Kron80}, also determined from the PSF-convolved $r$ band image). We measure  fluxes in all the bands for both regions. 
We define the ($NUV-r$) colour difference  $\Delta(NUV-r)$  as $(NUV-r)_{out}-(NUV-r)_{in}$ (likewise for  $\Delta(g-i)$), so that
negative values of $\Delta$ imply that the outer region 
of the galaxy is  {\em bluer} than the inner region\footnote{ We have checked that if we normalize the colour difference $\Delta(NUV-r)$ by the size of the galaxy,  the results presented in this paper do not change.}.

\begin{figure}
 \centering
\includegraphics[width=6cm]{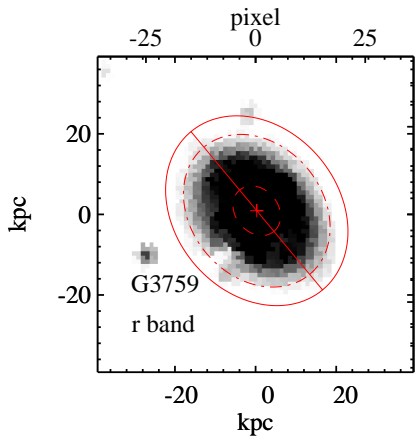}
\caption{The convolved $r$-band image of the GASS galaxy G3759. The dashed line shows the half-light ellipse, the dashed-dotted
line shows the 25 mag$/$arcsec$^2$ isophote, and the solid line shows the ellipse at  2.5 times the Kron radius.
The region within the dashed line is defined to be the ``inner'' region, while the region between the dashed and
the solid lines is defined as the ``outer'' region.}
 \label{fig:3759r}
\end{figure}

To illustrate,  Figure~\ref{fig:3759r} shows the HI-detected GASS galaxy G3759. The dashed curve indicates
the ellipse  
enclosing half the $r$-band light, the dot-dashed curve indicates the  25 mag/arcsec$^2$ isophote, and the 
solid curve the ellipse with a  semi-major axis equal to  2.5 times the Kron radius. 

\subsection{Image stacking}
\label{subsec:imagestacking}
In Appendix ~\ref{sec:testdnuvr}, we demonstrate that the 2-zone ($NUV-r$)
color measurement is reliable only when the radius of the inner zone is
larger than 6 arcsec (about the FWHM of the PSF of a GALEX NUV image)
and the error on the total (NUV-$r$) colour ($\sigma(NUV-r)$) is smaller than 0.1 mag.
If we were to apply both a size and an error cut, we would be left with   $\sim74\%$ of the
galaxies from the HI sample, $\sim36\%$ of the galaxies from the
C$_{M*}$ sample, $\sim70\%$ of the galaxies from the 
C$_{M*,NUV-r}$ sample and $\sim74\%$ of the galaxies from the 
C$_{M*,NUV-r,\mu*}$ sample. If we only apply the cut on size,   
$78\%$ of the HI sample, $58\%$ of the C$_{M*}$ sample,
$76\%$ of the C$_{M*,NUV-r}$ sample and $80\%$ of the C$_{M*,NUV-r,\mu*}$ sample remain. 

One problem with making a cut in $\sigma(NUV-r)$ is that the remaining  galaxies  are
biased toward NUV-bright (i.e. blue) objects, especially in
the $C_{M*}$ sample. In order to avoid this problem, we cut the  sample only by size, and
we stack the images in bins of stellar mass and HI gas fraction in order to boost the $S/N$ of the colour measurements. 
We then measure average colour and SFR gradients for the stacked images.
The increased $S/N$ of 
the stacked images will also improve the quality of the SED
fitting (Section~\ref{subsec:SF_tech}).

We first select  galaxies which have $R_{50}$ larger than 6$''$,
from the HI and ``parent'' samples. We then make new 
C$_{M*}$, C$_{M*,NUV-r}$ and  C$_{M*,NUV-r,
\mu*}$ control samples 
as described in
Section~\ref{subsec:controlsample}. Because the ``galaxy pool''
from which we draw these control galaxies is now smaller, each galaxy from the HI sample
only has one control galaxy matched in stellar mass and one control
galaxy matched in stellar mass and $(NUV-r)$ colour. For simplicity, we
will still call these three new samples the HI sample, 
C$_{M*}$ sample, C$_{M*,NUV-r}$ sample and C$_{M*,NUV-r,
\mu*}$ sample.

We divide the new  HI sample into 4 stellar mass bins with log
$M_*/M_{\odot}$ ranges of 10-10.25, 10.25-10.5, 10.5-10.75 and
10.75-11.5. Each stellar mass bin includes 80-90 galaxies. In each stellar mass bin, we further divide the galaxies
into two groups at  the median value of M(HI)/M$_*$. In total, the HI sample is divided  into 8 groups for stacking, with 40-50 galaxies in each group. We
also divide the matched C$_{M*}$ and  C$_{M*,NUV-r}$ samples
into 8 groups for stacking; each group in these control samples will then be matched to
the corresponding group from  
the  HI sample.

Our stacking procedure consists of the following steps. We rescale the images 
to the median value of $R_{50}$ for the galaxies in the group.  We
subtract the background, correct for Galactic extinction 
and any offset in  photometric zeropoint,  
align the centers of  all the images,  and then add them together. We
create stacked images for the images (convolved to the resolution of
the NUV image) for all 7 photometric bands. 
The stacked $r$-band image is used  as the
reference image to measure photometric parameters. 
We note that the stacked images all have a $\sigma(NUV-r)$
less than 0.01. 

\subsection{2-zone specific star formation rates}
\label{subsec:SF_tech}

We have used a spectral energy distribution (SED) fitting 
technique to  derive the average specific star formation rates (sSFR) in the inner and
outer regions of our stacked sample of galaxies.    
We follow the method from \citet{Salim07}, hereafter S07, with a few key changes that are detailed below. 
S07 used the \citet{BC03} population synthesis code to create a library of 100,000 model SEDs 
in 7 bands (FUV, NUV, $u, g, r, i, z$,) by generating model galaxies with a  range of ages,  star formation 
histories, metallicities and dust attenuation strengths. 
For a given galaxy, they evaluated the $\chi^2$ goodness-of-fit by comparing
the observed photometry with each model, determined the relative weight for each model, 
and finally built a probability distribution function (PDF) for the model parameters.   
The average of the PDF is adopted as the a nominal estimate of the parameters. 

 S07 show that the derived  dust attenuation (the optical depth $\tau_V$) is sensitive to the 
assumed prior distribution of $\tau_V$ in the model library, so the prior distribution should be as realistic as possible. 
To solve this problem, we estimate approximate values of the attenuation $\tau_V(gas)$ from the Balmer decrement F(H$\alpha$)/F(H$\beta$) within the SDSS fiber and the \citet{Calzetti00} extinction curve, when both F(H$\alpha$) and F(H$\beta$) have $S/N>3$. When either F(H$\alpha$) or F(H$\beta$) has $S/N<3$, we adopt $\tau_V(star, fiber)/0.44$ from the measurements of the attenuation of the
stellar continuum of the galaxy provided in the  MPA$\slash$JHU catalog ({\it tauv\_cont}). 
This parameter is obtained by fitting 
Bruzual \& Charlot (2003) population synthesis models to the stellar continuum; 
the reddening may then be estimated by determining the extra ``tilt'' that
must be applied to the models in order to fit the shape of the observed spectrum.   

The prior distribution of $\tau_V(gas)$ and $\mu$ ($\tau_V(star)=\tau_V(gas)\times\mu$) values we
adopt in our model library is tuned to reproduce the $\tau_V(star, fiber)$
predictions for the whole HI sample. We adopt for $\tau_V(gas)$ a Gaussian distribution peaked at 1.78, with a width $\sigma=0.55$, and we adopt for $\mu$ a Gaussian ditribution peaked at 0.44  \citep{Calzetti00}, with a width $\sigma=0.4$. The Gaussian distribution for $\tau_V(gas)$ is trimmed so that it only spans values between 0 and 4, and the Gaussian distribution for $\mu$ is trimmed so that it only spans values between 0 and 1. Here we assume that the $\tau_V(star)$ does not change significantly throughout the galaxy.

In reality, the dust distribution in galaxies is more complicated 
than our simple model assumes. One issue is that for a 
typical galaxy, the relative attenuation between the inner disk, 
outer disk  and bulge depends strongly on 
inclination \citep{Tuffs04}. However, this effect becomes 
most notable when the inclination is large  (axis ratio $b/a<0.4$), 
so this problem 
is much mitigated by excluding galaxies with 
$b/a<0.4$ from the samples 
(Section~\ref{subsec:PSsample}) \citep{Yip10}. In the absence
of more detailed information on the dust distribution in each galaxy, 
we believe our method gives results that are as accurate as possible.

To test the robustness of the sSFR derived from our SED fitting, 
we also measured sSFR directly from the SDSS fiber spectrum,
and we compared this
to the sSFR estimated within the 3 $''$ aperture using our
SED fitting technique. The  SFR derived from
the spectrum is calculated  from the H$\alpha$ luminosity, 
and corrected for dust using the Balmer decrement F(H$\alpha$)/F(H$\beta$) 
and the \citet{Calzetti00} extinction curve. 
The stellar mass inside the SDSS fiber is taken from 
the MPA$\slash$JHU catalog. For SED fitting, we take the SDSS fluxes within the fiber from the SDSS archive (fibercounts), measure the GALEX NUV fluxes and convolved SDSS $u$ band and GALEX FUV fluxes within 6 $''$ around the galaxy (approximately the FWHM of the GALEX NUV PSF). Then we use the $u$ band fluxes (within the fiber aperture and from the convolved images) to normalize the GALEX FUV and NUV fluxes, so that we get consistent SED from the 7 bands within the 3$''$ aperture.

\begin{figure*}
 \centering
\includegraphics[width=12cm]{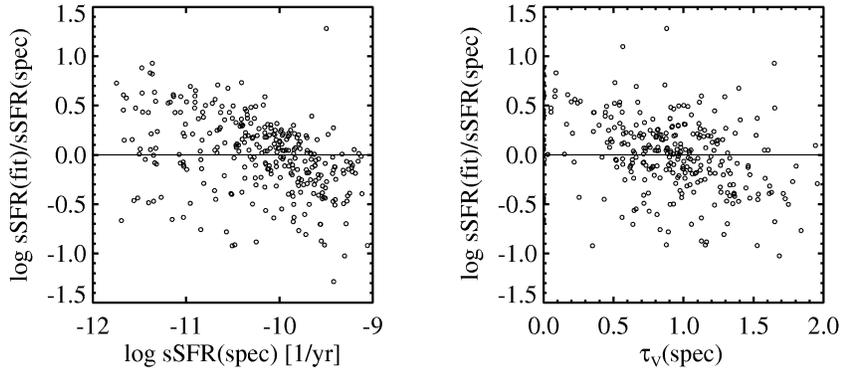}
 \caption{The difference between the sSFR derived from SED fitting
and the sSFR measured from the dust-corrected H$\alpha$ luminosity and stellar mass
from the SDSS fiber spectrum. The difference is plotted as a function of sSFR (left) and $\tau_V(gas)$ within the fiber (right).}
 \label{fig:fibeffect}
\end{figure*}

We see from Figure~\ref{fig:fibeffect} that the two estimates 
generally agree with each other, with a  1$\sigma$ scatter of
$\sim$0.39 dex. The offset between the two estimates 
exhibits weak systematic trends with both log SFR and $\tau_V$.
The most discrepant results are obtained for galaxies with 
very low specific star formation rates, where the UV luminosities are
low and may trace populations of stars that are not properly accounted 
for in standard population synthesis models (Conroy \& Gunn 2010).
We have also tested the effect of varying the adopted
priors for $\tau_V$ . We confirm that the prior that 
is most similar to the real distribution (as calculated from 
the spectrum), gives the best fit.  

\subsection{Morphological parameters}
\label{subsec:morph_tech}
We closely follow the procedure described  in \citet{Lots04}, hereafter L04, to derive the asymmetry ($A$) and smoothness ($S$)
parameters  from the SDSS $g$-band images. Both $A$ and $S$ are measured inside an aperture equal to 1.5 times 
the petrosian radius ($r_p$). All neighboring objects are masked. 

$A$ is a measure of the difference between a given galaxy image and the image rotated by 180 degrees 
about the object center (the center is determined by minimizing $A$). A higher value of $A$ 
means that the galaxy is more asymmetric. $S$ is a measure of the difference between 
a given galaxy image and the image  smoothed with a 0.2$r_p$ wide boxcar kernel.
A higher  value of $S$ implies that the galaxy has a more clumpy morphology on scales 
equal to the kernel size. $A$ and $S$ are also calculated using regions of blank sky in the
vicinity of the galaxy; the average values of $A$ and $S$ calculated for the background are subtracted from the
measurements we make for the galaxy.

\begin{figure*}
 \centering
 \includegraphics[width=12cm]{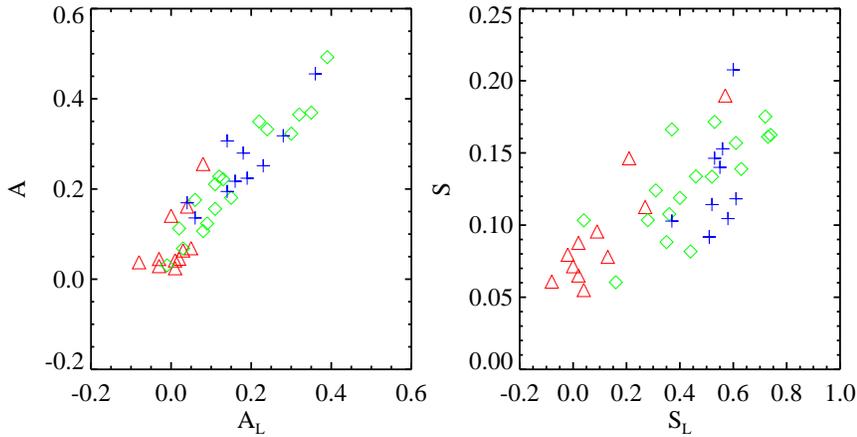}
 \caption{Comparison of morphological parameters measured by L04 (x-axis) and by us
(y-axis).  The red triangles are Sa galaxies, green diamonds are Sb galaxies and blue crosses are Sc galaxies.}
 \label{fig:as_lotz}
\end{figure*}

To make sure that our measurements are consistent with L04, we measure $A$ and $S$ for 41 galaxies from L04 
with images available from the SDSS. Our results are shown in Figure~\ref{fig:as_lotz}. 
We see that our parameters and those from L04 follow  similar trends along the Hubble sequence.
There is quite close agreement between the $A$ values, but our $S$ values are typically four times smaller than
those of Lotz -- this is likely to be a reflection of the different quality/resolution of the
images or different details in the image processing steps in the two cases. Nevertheless, our test shows that the relative trends in $S$ 
are the same for both sets of measurements, so these parameters are a useful diagnostic of 
relative changes in galaxy morphology.  

We visually check all the images in our sample using  the SDSS Imaging Finding 
Chart Tool \footnote[3]{http://cas.sdss.org/astro/en/tools/chart/chart.asp}, and
we identify galaxies that are clearly undergoing  a merger event. 
We find no enhancement in the merger fraction for the galaxies in the HI sample 
(9, 8, and 7 galaxies out of a total of 519 in the  HI, C$_{M*}$, and C$_{M*,nuvr}$  samples
exhibit clear signs of a merger or interaction in our images).

\section{Results}
\label{sec:results}

As shown in Figures 2 and 3,  galaxies from the HI sample are bluer, and have larger sizes and  
later type morphologies compared to  
C$_{M*}$ galaxies. This is not surprising, because it is well known that galaxies that are more actively 
star-forming also contain more gas. 
The C$_{M*,NUV-r}$ and C$_{M*,NUV-r,\mu*}$ galaxies, on the other hand, are matched to the HI sample both in stellar mass and in global NUV-$r$ colour, so if
HI gas fraction and star formation activity track each other very closely, one might expect the galaxies in these two
samples to have identical properties. In the following sections, we will show that this is not true.
We divide the HI sample into four stellar mass bins, and study how 
sizes and colour gradients  vary with  atomic gas mass fraction. 
We also compare  properties of the corresponding control galaxies along the same sequence.
The main  purpose of comparing our results with those derived from control samples 
is to isolate those trends that can be attributed to
increasing HI content, rather than to any other correlated property, 
such as stellar mass or global star formation rate. Because the samples have all been matched in redshift, the noise in the
 measurements of luminosity and colour necessary to estimate quantities
 such as stellar mass, will be identical in all the comparison samples.

\subsection{HI mass fraction and the optical sizes of galaxies}
\begin{figure*}
 \centering
 \includegraphics[width=12cm]{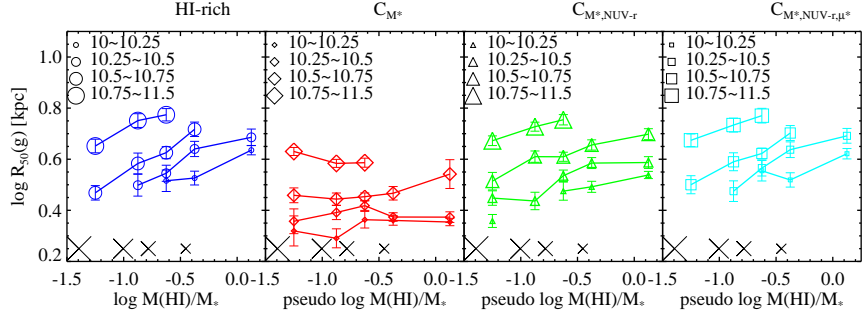}
 \caption{Relation between galaxy size (from the $g$ band) and HI mass fraction for galaxies in the HI sample (blue circles), C$_{M*}$ control sample (red diamonds), C$_{M*,NUV-r}$ control sample (green triangles) and C$_{M*,nuvr,\mu*}$ control sample (cyan squares).
The control galaxies have no HI mass measurements, and are just plotted
in the same M(HI)$/$M$_*$ bin as the corresponding galaxies in the HI sample. Symbols of different sizes are for different  stellar masses as indicated
 at the corners of the plots. The error bars show the
r.m.s. deviation in the mean galaxy size calculated
through bootstrapping. The crosses at the bottom of the plot
mark the location of  the mean gas fraction for galaxies  in each stellar mass bin
as given in  C10.}
 \label{fig:size4bands}
\end{figure*}

Figures~\ref{fig:quality_lgm} and \ref{fig:quality_lgm_nuvr} demonstrate that the galaxies in the HI  sample 
have larger average sizes than both $C_{M*}$ and $C_{M*,NUV-r}$ control sample galaxies. The mean value of $R_{50}(i)$ of the HI sample is 
larger by 1.2 and 0.3 kpc than the mean values of $R_{50}(i)$ of the $C_{M*}$ and $C_{M*,NUV-r}$ control sample galaxies respectively.
In  Figure~\ref{fig:size4bands}, we plot the relations between the half-light radius $R_{50}$ and M(HI)$/$M$_*$ in
four different bins of stellar mass for galaxies in the HI sample. 
The mean values of M(HI)$/$M$_*$ for each stellar mass bin 
are  marked as  crosses of different sizes at
the bottom of the plot. These have been derived by Catinella et al (2010) 
using GASS survey galaxies for which HI masses
have been measured down to a limiting M(HI)$/$M$_*$ limit of $\sim 1.5 \%$.   
The majority of  our HI-detected galaxies have atomic gas fractions that are above the average value --
this is not surprising since galaxies detected in the ALFALFA survey make up the 
bulk of our sample. 

We  see that the half-light radius R$_{50}$ measured in the $g$-band 
increases as a function of  M(HI)$/$M$_*$ at a given value of $M_*$. 
This result is consistent with the scaling relations published in Zhang et al (2009) and  C10, 
which clearly showed that
M(HI)$/$M$_*$ correlates strongly with  stellar surface density $\mu_*$   
(i.e. with galaxy size at a fixed value of the stellar mass). Indeed, both studies  find that
M(HI)$/$M$_*$  can be best predicted using a  {\em combination} of colour and  $\mu_*$.
As discussed in  Zhang et al (2009), these scalings can be understood in terms of
the  Kennicutt-Schmidt law of star formation \citep{Kennicutt98}, which states that the surface density of
star formation scales with the surface density of cold gas as a power-law with slope
$\sim 1.4$. A star formation law of this form leads to the expectation that 
\begin {equation} \log M(HI)/M*= a \log \mu_* + b \log SFR/M_* +c, \end {equation}
where $a,b$ and $c$ are constants. The NUV-$r$ colour is an excellent proxy for
SFR/$M_*$ (especially in the blue sequence), so this leads to a prediction of a linear relation linking
HI mass fraction, NUV-$r$ colour and stellar surface mass density. 
The two right hand panels of Figure~\ref{fig:size4bands} show that sizes of  $C_{M*,NUV-r}$
and  $C_{M_*,NUV-r,\mu_*}$ control galaxies
exhibit the same  increase as for the  HI-detected galaxies. This  supports our hypothesis that
the scaling of galaxy size with HI mass fraction arises as a {\em consequence} of the star formation
rate-gas surface density relation.

\begin{figure*}
 \centering
 \includegraphics[width=12cm]{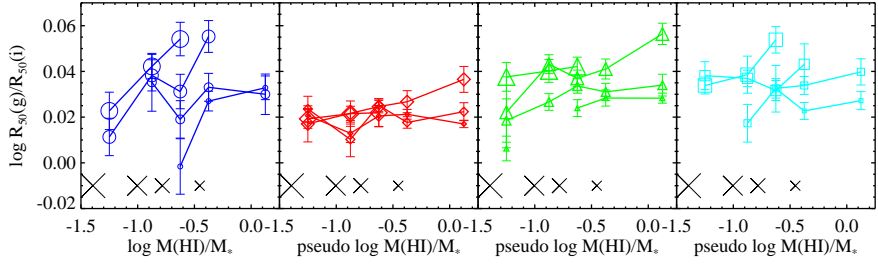}
 \caption{As in Figure 8, except for the  ratio between the $g$ and $i$-band
half-light radii.}
 \label{fig:sizeratio4bands}
\end{figure*}

In  Figure~\ref{fig:sizeratio4bands}, we analyze trends in the ratio 
of the $g$-band and $i$-band half-light radii as a function
of HI mass fraction.
R50 defined in the $g$-band is always slightly larger than R50 defined in the $i$-band, because 
the light from younger stellar populations is generally spread over a larger effective radius
than the light from older stellar populations.
The ratio between the $g$ and $i$-band  radii increases  
as a function of gas fraction, and is also larger for more massive galaxies at a fixed value of
M(HI)/$M_*$. As we will show in the next section, this difference in $g$- and $i$-band half-light radii is also
found when we analyze  $g-i$ colour gradients. In the  panels to the right, we plot
size differences for the three control samples.  Neither the 
$C_{M*}$ nor the $C_{M*,NUV-r}$ control samples show  
similar trends to the HI sample.
The size differences do increase for more massive galaxies
in  the $C_{M*,NUV-r,\mu_*}$ control sample, but the effects are weaker than
those observed for the sample with real HI mass measurements. 

One might worry whether these 
trends arise because HI-rich galaxies are more dusty. We analyzed the correlation between  optical depth $\tau_V$ 
and HI mass fraction, and find that $\tau_V$ does not increase with $M(HI)/M_*$.
In addition, control galaxies matched in stellar mass and colour do not have smaller $\tau_V$ values 
than  galaxies from the HI sample. 
We are thus led to the conclusion that there 
is an {\it intrinsic} correlation between
M(HI)/M$_*$ and R50(g)/R50(i).  

\subsection{HI mass fraction and ($g-i$) colour gradients}

\begin{figure*}
 \centering
 \includegraphics[width=12cm]{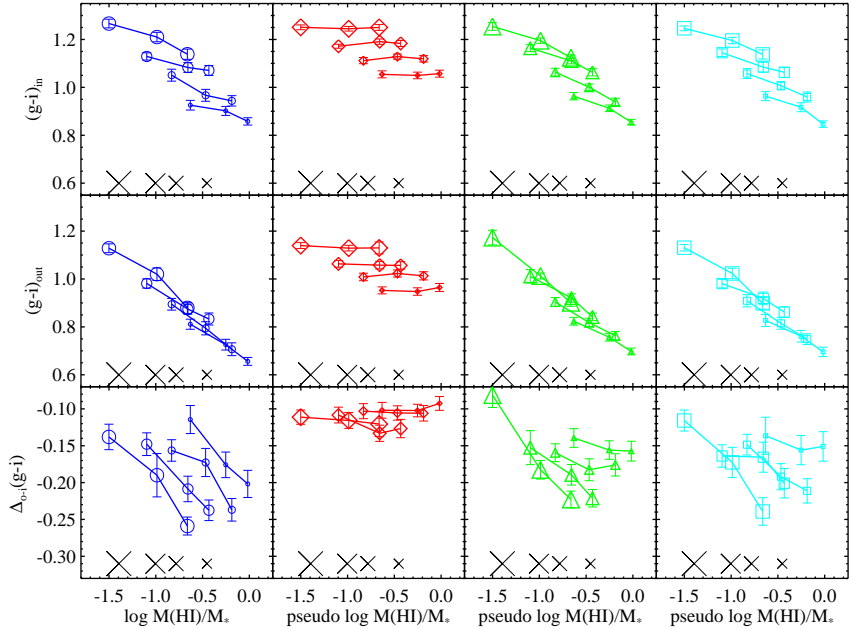}
 \caption{Relation between inner $g-i$ colour, outer $g-i$ colour and $\Delta_{o-i}(g-i)$ and HI mass fraction for the HI sample (left column, plotted as blue circles),
C$_M*$ control galaxies (second column, plotted as red diamonds)
, the C$_{M*,NUV-r}$ control galaxies (third column, plotted as green triangles) and the C$_{M*,NUV-r,\mu*}$ control galaxies (fourth column, plotted as cyan squares).
The control galaxies  have no HI mass measurements, and are plotted
in the same M(HI)$/$M$_*$ bin as the corresponding
galaxies in the HI sample. The sample is divided into 4
stellar mass bins, and plotted with symbols of different sizes
as indicated in Figure~\ref{fig:size4bands}. The error bars show the r.m.s. deviation
in the mean values calculated through bootstrapping.
The crosses denote the mean gas fraction for galaxies in each
stellar mass bin (from C10).}
 \label{fig:gi2z}
\end{figure*}

In the top two panels of  Figure~\ref{fig:gi2z},  we analyze trends in the  inner and outer $g-i$ colours of the galaxies
in our four samples. In the bottom panel, we look at trends in $(g-i)$  colour gradients. Our results are as follows:

{\bf Inner colours:} The inner $g-i$ colour of a galaxy depends strongly on stellar mass, in the sense
that  more massive galaxies have redder inner colours.
The inner $g-i$ colour depends  only weakly on HI mass fraction at a given value of $M_*$. 
We see very similar trends in inner $g-i$ colour 
for control samples matched in $NUV-r$ colours.
We conclude  that  the inner
colour of a galaxy is  primarily sensitive to its stellar mass.

{\bf Outer colours:} The behaviour of  outer $g-i$ colour as a function of stellar mass and
HI fraction is very different to that of the inner colour. It depends strongly on  
HI mass fraction, in the sense that more HI-rich galaxies have bluer outer colours.
At a fixed value of the HI mass fraction, there is no dependence on stellar mass.
We see very similar trends in the $C_{M_*,NUV-r}$ and
$C_{M_*,NUV-r, \mu_*}$ control samples, but not
in the $C_{M_*}$ control sample. This indicates that the outer colour of a galaxy
is primarily sensitive to how much ongoing star formation there is in the galaxy.  

{\bf Colour gradients:} The $g-i$ colour gradient of a galaxy depends 
strongly on its HI mass fraction, in the sense that more HI-rich galaxies
become bluer on the outside relative to the inside. At a fixed value of
M(HI)/M$_*$, more massive galaxies have stronger colour gradients.
 None  of the control samples exhibits exactly the same trends as  the HI sample.
In the  $C_{M_*,NUV-r}$ and $C_{M_*,NUV-r,\mu_*}$  samples, the colour gradients 
for massive galaxies  behave in much the same way as for the sample with HI measurements.
For low mass galaxies, however, there is no trend in colour gradient with
log M(HI)/$M_*$ for the control galaxies.
Our main conclusion, therefore, 
is that there is an intrinsic link between higher HI fractions and bluer outer disks.

\subsection{HI mass fraction and gradients in  $(NUV-r)$ colour and sSFR}
\label{subsec:nuvr_result}

In this section we analyze  trends in the inner and outer NUV-$r$ colours and
colour gradients as a function of HI mass fraction. The analysis is similar to that of the $g-i$ colours in the previous 
section, except that in this case, the average inner and outer colours
are computed from stacked images due to the lower $S/N$ of the NUV images. As we will show, the main advantage of studying
NUV-$r$ colours is the fact that the             
UV is a much more sensitive 
probe of the presence of young stars than the $g$-band, so that both the trends as a function of $M_*$
and M(HI)/$M_*$,  and the
differences between the HI sample and the control samples found in the previous section are much
enhanced.

\begin{figure*}
 \centering
 \includegraphics[width=12cm]{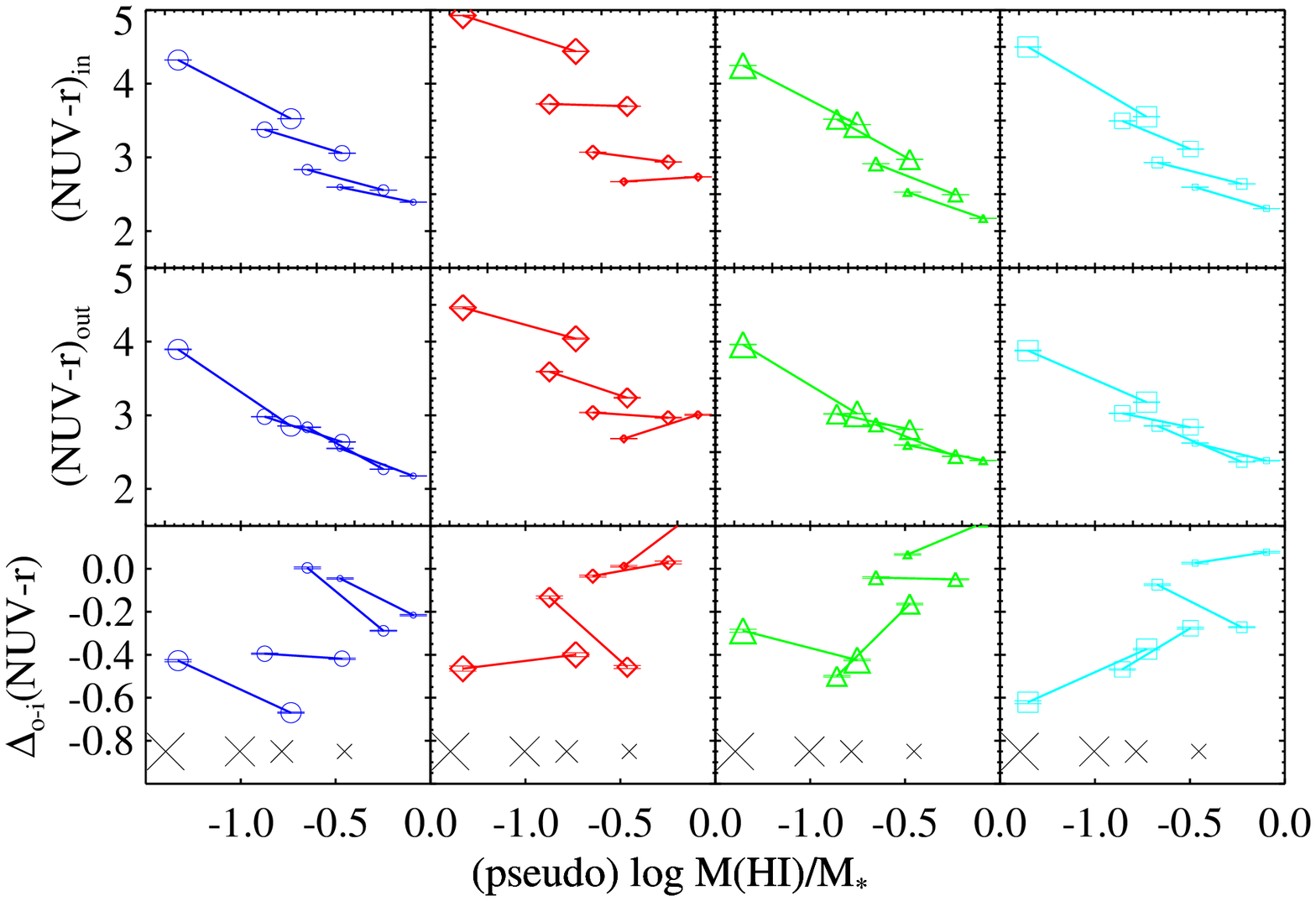}
 \caption{Relation between inner $NUV-r$, outer $NUV-r$, $\Delta_{o-i}(NUV-r)$ and HI mass fraction for the galaxies from the HI (blue circles) sample, and C$_{M*}$ (red diamonds), C$_{M*,nuvr}$ (green triangles) and C$_{M*,nuvr,\mu*}$ (cyan squares) galaxies. The colours are measured from the stacked images. Note that the control galaxies have no HI mass measurements, and are plotted in the same M(HI)$/$M$_*$ bin as the corresponding galaxies in the HI sample. Symbols of different sizes indicate different stellar masses as denoted in Figure~\ref{fig:size4bands} (larger symbol corresponds to higher stellar masses). The crosses denote the mean gas fraction for galaxies in each
stellar mass bin (from C10).}
 \label{fig:stackdnuvr2d}
\end{figure*}

The results shown in  Figure~\ref{fig:stackdnuvr2d} confirm essentially all
the conclusions discussed in the previous section. The main qualitative difference we
find by studying NUV-$r$ colours rather than $g-i$ colours is that the behaviour of 
the colour gradients and the outer NUV-$r$ colours as a function of HI mass
fraction now differ more significantly  from
those exhibited by both the C$_{M*,NUV-r}$ and the 
C$_{M*,NUV-r,\mu_*}$ control galaxies. This strengthens our
conclusion that the trend in the colour of the outer disk is fundamentally linked to
the increase in the HI content of the galaxy.   

\begin{figure*}
 \centering
 \includegraphics[width=15cm]{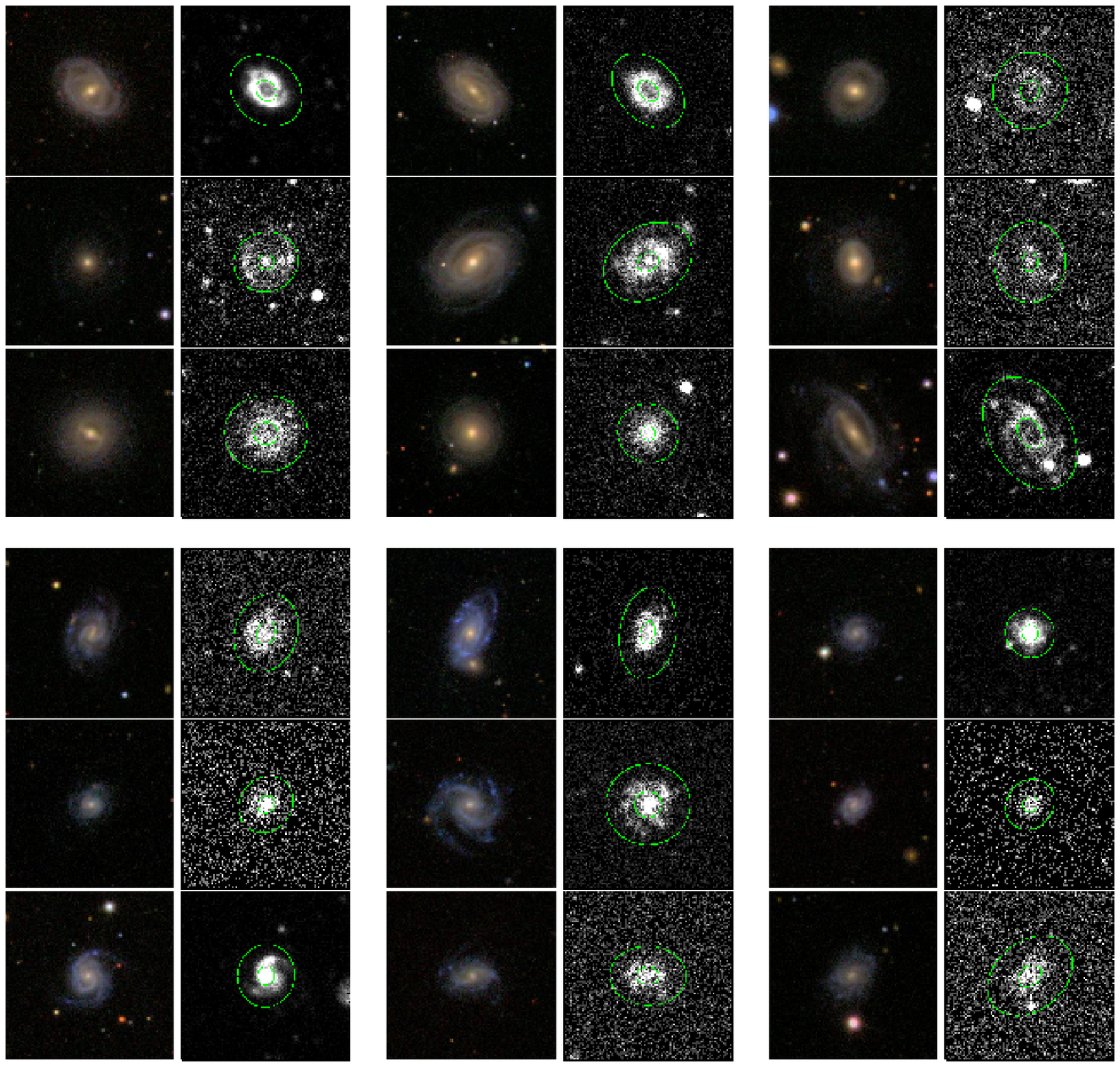}
 \caption{SDSS and GALEX NUV images of galaxies from the HI sample which
   have the  most negative ($NUV-r$) colour differences
   ($\Delta_{o-i}(NUV-r)$<-1.1) (the first 3 rows), and of galaxies that
   have the highest  HI mass fractions (M(HI)$/$M$_*$) ($\sim1$)
   (last 3 rows). For each image pair, the left image is the SDSS
   $g,r,i$ colour composite image, and the right image  is the NUV map
   for the same galaxy.}
 \label{fig:dnuvr_example}
\end{figure*}

We note that Figures ~\ref{fig:gi2z} and ~\ref{fig:stackdnuvr2d} both show that it is
the most massive galaxies with stellar masses $\sim 10^{11}$ M$_{\odot}$, and with
relatively low  HI gas fractions of $\sim$ 10\%, that have 
the strongest colour gradients.
Low mass galaxies have the highest HI gas fractions, but  their colour gradients are
relatively weak. It is interesting to investigate the morphology of the UV light
for both kinds of systems.
Figure~\ref{fig:dnuvr_example} shows optical and NUV 
images for the 18 HI-sample galaxies
with  the most negative
$\Delta_{o-i}(NUV-r)<-1.1$ as well as the 18 HI-sample galaxies which have the
highest values of  $M(HI)/M_*$ in the HI sample (we point out that, because the HI sample is stellar mass limited at
10$^{10}$ M$_{\odot}$, we are missing a population with the highest HI mass fractions ($M(HI)/M_*$) at a fixed M(HI)
). 
The galaxies with the most negative $\Delta_{o-i}(NUV-r)$ all have
very red central regions and a blue outer disk. Many of them have bars, and the blue
light tends to be located in a UV-bright ring around the bar (upper panels of Figure~\ref{fig:dnuvr_example}).
Galaxies with the highest HI gas fractions 
are usually blue throughout the whole galaxy  (lower panels of Figure~\ref{fig:dnuvr_example}). 

\begin{figure*}
 \centering
 \includegraphics[width=12cm]{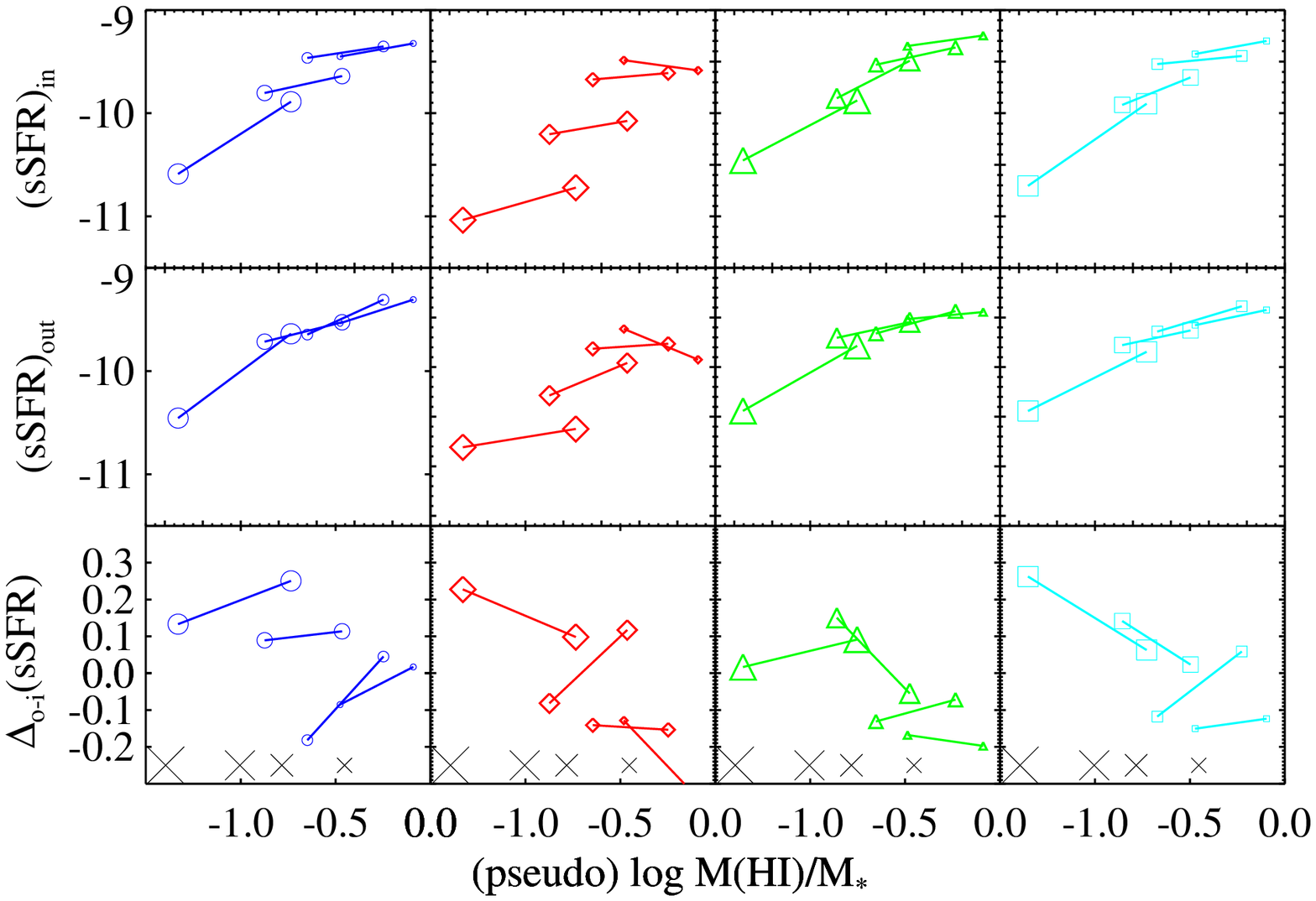}
 \caption{Relation between inner $sSFR$, outer $sSFR$, $\Delta_{o-i}(sSFR)$ 
and HI mass fraction for galaxies 
from the HI (blue circles) sample, and from the  C$_{M*}$ (red diamonds), 
C$_{M*,nuvr}$ (green triangles) and  C$_{M*,nuvr,\mu*}$ (cyan squares) control
samples. The sSFR are derived from SED fitting as explained in the text, 
and the SEDs are measured from the stacked images. 
Note that the control galaxies have no HI mass measurements, 
and are plotted in the same M(HI)$/$M$_*$ bin as the 
corresponding galaxies in the HI sample. Symbols of 
different sizes indicate different stellar masses 
as denoted in Figure~\ref{fig:size4bands} 
(larger symbol corresponds to higher stellar masses). The crosses denote the mean gas fraction for galaxies in each
stellar mass bin (from C10).}
 \label{fig:stackdssfr2d}
\end{figure*}

Finally, Figure~\ref{fig:stackdssfr2d} shows  plots equivalent to 
Figures ~\ref{fig:gi2z} and  ~\ref{fig:stackdnuvr2d} for the specific star formation
rates we  derive from  fitting models to the UV through optical SEDs 
derived from the stacked images (see section 3.3). We caution that
we have to make assumptions about the likely range of
dust attenuation in our galaxy sample, as well as assumptions about
how the dust might be distributed throughout the galaxy.
We thus regard these results as indicative, rather than definitive.
The main point we wish to make with this plot is that with {\em plausible
assumptions} about extinction, we find that outer disks are younger, with
a higher ratio of present-to-past averaged star formation in 
galaxies with higher HI mass fractions. The same effect is not seen as clearly or strongly in 
the control samples. This implies that the outer disks of galaxies with higher HI mass fractions are 
are growing more rapidly, i.e. that these galaxies are forming from the "inside-out".
 
We have estimated the uncertainty in our estimated sSFR arising from the
age-metallicity degeneracy. The offset
between the estimated value of the sSFR will be less than 0.2 dex for galaxies with half-solar and
twice solar metallicities. Note also, that all the galaxies in this analysis 
have $M_*> 10^{10} M_{\odot}$;  the mass-metallicity relation \citep{Gallazzi05}
implies that there will be no very low metallicity galaxies included in our analysis.
 
\subsection{HI gas fraction and higher order morphological parameters}
\label{subsec:morph_result}
\begin{figure*}
 \centering
 \includegraphics[width=12cm]{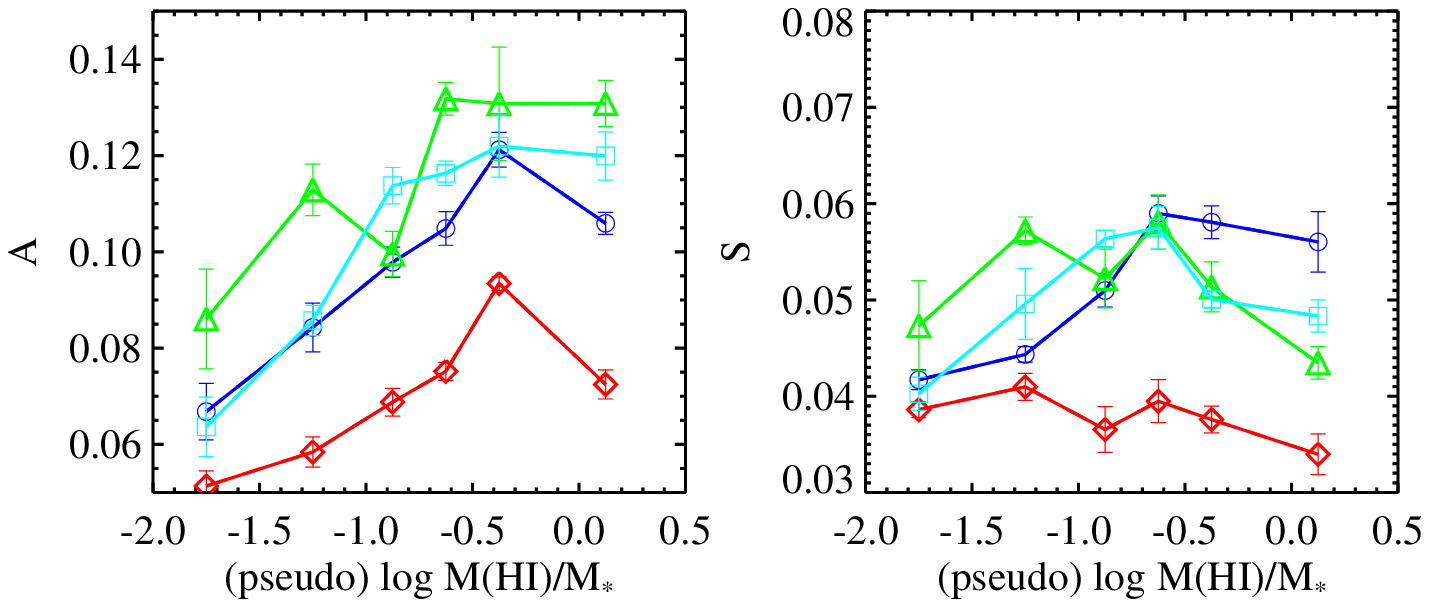}
 \caption{Relation between the median values of the 
morphological parameters asymmetry ($A$) and smoothness ($S$) and
HI mass fraction for galaxies in the HI sample (blue symbols and lines). 
The median $A$ and $S$
values  of the corresponding $C_{M*}$ galaxies
are plotted in red; results for  $C_{M*,NUV-r}$ galaxies are 
shown in green; results for $C_{M*,NUV-r,\mu*}$ galaxies are shown in cyan.
A median value is plotted only when there are more 
than 6 galaxies in that bin.  Error bars show the
r.m.s.  deviation calculated through bootstrapping.}
 \label{fig:morph_HI}
\end{figure*}

\begin{figure*}
 \centering
 \includegraphics[width=12cm]{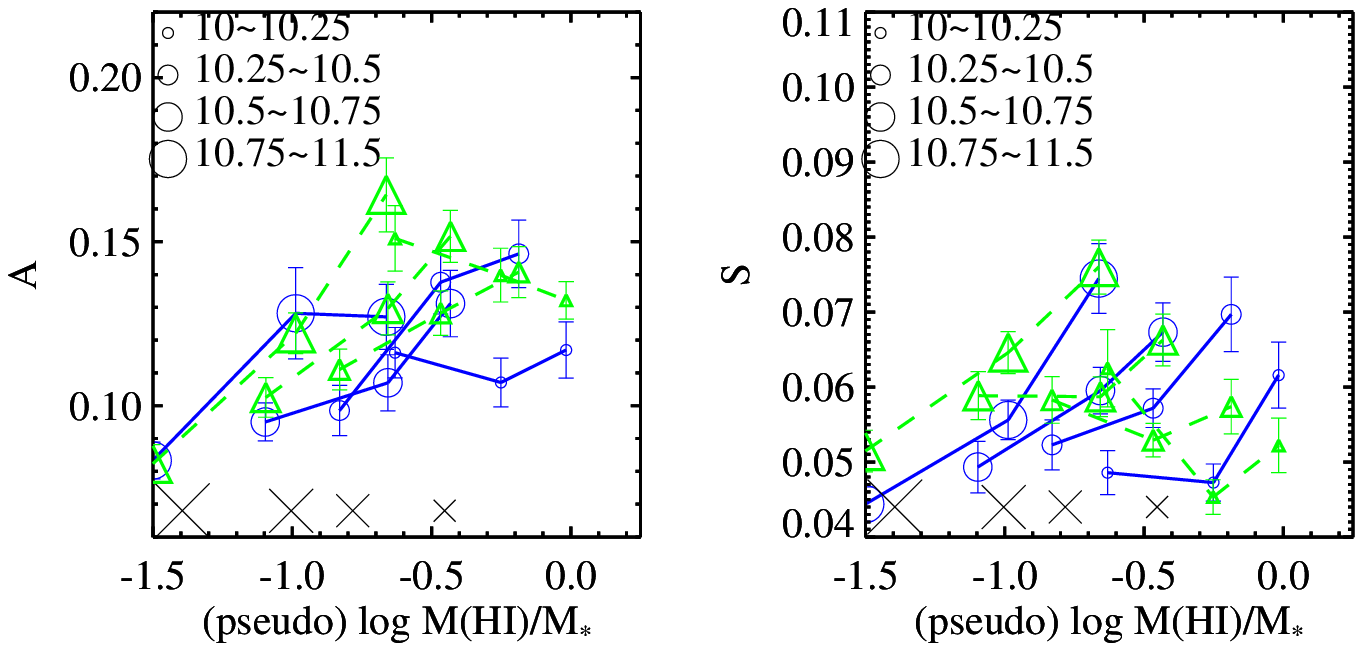}
 \caption{Relation between $A$ and $S$ and HI mass fraction for galaxies 
from the HI (blue circles) sample and from the  C$_{M*,NUV-r}$ (green triangles) control sample. The sample is divided into 4
stellar mass bins, indicated by symbols of different sizes as described in the corner
of each panel. The error bars show the r.m.s. deviations in the mean values
of $A$ and $S$  calculated through bootstrapping. The crosses denote the mean gas fraction for galaxies in each
stellar mass bin (from C10).}
 \label{fig:morph_lgm_frac}
\end{figure*}

Figure~\ref{fig:morph_HI} shows $A$ and $S$ as a function of HI 
fraction for the galaxies in the HI sample (blue), and for galaxies in the 
$C_{M*}$ (red),  $C_{M*,NUV-r}$ (green) and $C_{M*,NUV-r, \mu_*}$ (cyan) 
control samples. Galaxies in the HI sample are significantly 
 more asymmetric and less smooth 
than galaxies in the $C_{M*}$ sample. There is no significant  difference between the  
HI,   $C_{M*,NUV-r}$ and  $C_{M*,NUV-r,\mu_*}$ samples.
In Figure~\ref{fig:morph_lgm_frac}, we plot $A$ and $S$ as a function of HI mass fraction for
galaxies in  the HI sample (blue) and in the  $C_{M*,NUV-r}$ control sample (green) for 
 different stellar mass bins. We see that at fixed stellar mass,  $A$ and $S$
both increase as a function of   M(HI)/M$_*$ for galaxies in the HI sample. At fixed stellar mass,
galaxies in $C_{M*,NUV-r}$ sample have similar
values of the asymetry parameter $A$ and smoothness parameter $S$ (except at the very highest HI mass fractions).  

We conclude that the correlation of galaxy asymmetry and smoothness with
HI mass fraction arises {\em because} galaxies with higher HI gas fractions have younger stellar populations.
Reichard et al (2009) show that there is a strong link between galaxy asymmetry (in their case, as measured by
the  $m=1$ azimuthal Fourier mode) and the  age of its stellar population. They found that this
link was independent of the other structural properties of the galaxy. Our analysis of the control samples 
appears to indicate that this link is also independent of HI content.   
Lopsidedness/asymmetry is believed to trace a non-equilibrium dynamical state caused by 
mergers, tidal interactions, asymmetric accretion of gas, or asymmetries related to the dark matter halo \citep{Bournard05}.   
As discussed in section 3.4, we found no enhancement in the fraction of galaxies
undergoing mergers in the HI sample relative to control samples, so it is perhaps not too surprising that we find no intrinsic correlation between lopsidedness/asymmetry and HI content.

\section{ Discussion and Conclusions} \label{sec:summary} 
We have  analyzed a sample
of galaxies from the $parent$ $sample$ of the GASS survey, for which HI
mass measurements are available  from the GASS survey itself and from
ALFALFA. At a given value of $M_*$, our sample consists primarily of
galaxies with larger-than-average values of M(HI)/$M_*$ .
We use a combination of optical  photometry from the SDSS and FUV,NUV
images from GALEX to study properties of the 2-zone stellar populations
of these HI-rich systems, and we investigate how they vary with HI
mass fraction.

We constructed three control samples for comparison with the galaxies
with HI mass measurements: one sample is  matched in  stellar mass
and redshift (C$_{M*}$),  the second one is matched in  stellar
mass, $NUV-r$ colour and redshift (C$_{M*,NUV-r}$) and the third one is matched in  stellar
mass, $NUV-r$ colour, surface mass density and redshift (C$_{M*,NUV-r}$).  We then generated
self-consistent 7-band photometry  (FUV, NUV, $u$,$g$,$r$,$i$ and $z$)
for the galaxies in all three samples and derived inner
colours, outer colours, asymmetry and smoothness parameters for all the
galaxies. We also generated four sets of stacked images in different
bins of stellar mass and HI gas fraction, and used these to derive inner
and outer specific star formation rates using an SED-fitting technique.

The main  purpose of comparing our results with those derived from
control samples is to isolate those trends that can be attributed to
increasing HI content, rather than to any other correlated property,
such as stellar mass or global star formation rate. As  expected, HI-rich
galaxies differ in nearly all their properties from galaxies of the
same stellar mass  selected without regard
to HI content.  Most  of these
differences are  attributable to the fact that galaxies with more
gas are bluer and more actively star-forming. 
In order to identify those  galaxy properties that are {\em truly causally connected
with HI content}, we also compare our
results derived for the HI sample with those derived for galaxies with
the same distribution of both stellar mass {\em and} NUV-$r$ colour. 
Finally, Catinella
et al (2010) have  demonstrated that the best {\em predictor} of the HI gas
fraction is a linear combination of NUV-$r$ colour and the logarithm
of the stellar surface mass density of the galaxy. At a fixed colour, galaxies
with lower stellar mass densities have higher HI gas fractions.  
This motivates comparison with a control sample that is matched in stellar mass,
NUV-$r$ colour {\em and} stellar surface mass density.

We find that the only photometric property  that is clearly
attributable to increasing HI content and not to the total amount of star
formation in the galaxy, is the {\em colour gradient} of the galaxy.
There appears to be a causal link between HI fraction  and outer
disks that are bluer with respect to the inner part of the galaxy. The
same effect also manifests itself as a systematic increase in the ratio
of size of the galaxy measured from SDSS images in the  $g$-band to that 
measured  in the $i$-band. 
The colour gradients are significantly stronger if one analyzes UV/optical colours
than if one uses $g-i$ as a tracer.  Our SED modelling demonstrates that
as the HI fractions of galaxies increase, the specific star formation
rates in their outer regions become higher with respect to their inner
regions. This means that the outer regions of HI-rich galaxies are
{\em younger}. 

Control samples that are matched in size in addition 
to colour and stellar mass, tend to exhibit stronger colour gradients, but
the observed scalings are still weaker than for the sample with real HI
mass measurements. This indicates that  colour gradients and size are
partially co-variant, as might be expected if infalling gas was recently
added to the outside of the disk in the most HI-rich galaxies in our sample.  

The ``inside-out'' picture of disk galaxy formation has commonly served
as a basis for semi-analytic  modelling the chemical evolution of disk
galaxies (e.g. Chiappini et al. 1997), as well as the formation of disk
galaxies in the context of   Cold Dark Matter cosmologies (e.g. Kauffmann
1996; Van den Bosch 1998; Boissier \& Prantzos 1999; Dutton 2009).
In this scheme, gas in the halo surrounding the  galaxy cools, falls
onto the galaxy while conserving its angular momentum,
and fuels star formation in the disk. Gas accreted at  late time has
higher specific angular momentum, and so settles in the outer region
of the galaxy.   
However, hydrodynamical simulations of the formation
of disk galaxies indicate that angular momentum is not conserved and
disks do not always form from the inside out (Sommer-Larsen et al 2003;
Robertson et al 2004).

Direct observational evidence  that  disk galaxies  form from the inside
out has thus far been quite difficult to establish.  Pohlen \& Trujillo 
(2006) found that 60\% of local disk galaxies have inner exponential
profiles followed by a steeper (downward bending) outer exponential
profile, while 30\% have a shallower (upward bending) outer profiles.
Azzollini et al (2008) find that for a given stellar mass, the radial
position of this break has increased with cosmic time by a factor of
1.3+/-0.1 between $z=1$ and 0, in agreement with a moderate inside-out
growth of disk galaxies over the last $\sim  8$ Gyr. Cresci et al
(2009) find evidence for evolution in the zero point of the Tully-Fisher
relation of galaxy disks at $z \sim 2$ with respect to local samples that
is in agreement with cosmological simulations of disk galaxy formation.

Coherent studies of the   metallicity, colour and star formation rate
gradients in local galaxies have been  few and far between.
Zaritsky, Kennicutt \& Huchra (1994) performed a careful study of
abundance gradients derived from HII regions  in a sample of 53 nearby
spiral galaxies. They found that the slopes of the radial abundance
gradients when rescaled to units of dex/isophotal radius, do not
correlate with the luminosity or Hubble type of the galaxy. They did
find evidence that galaxies with bars have flatter abundance gradients,
which suggests that gas inflows may be  important in  understanding
the time evolution of metallicity gradients. \citet{deJong96} investigated the effects of star formation history and the metallicity on the colour profiles of 86 face-on spiral galaxies, and found that the outer regions of disks are on average younger and of lower metallicity. Munoz-Mateos et al (2007)
studied specific star formation rate profiles of a sample of 160  nearby
spiral galaxies from the GALEX Atlas of Nearby galaxies \citep{Gildepaz07}. They found
a large dispersion in the slope of the sSFR profiles, with a slightly
positive mean value, which they interpreted as implying  moderate net
inside-out disk formation. They found that in a minority of galaxies,
the scale length of the UV light is significantly larger than the scale
length of the near-IR light, and interpreted these galaxies as having
undergone an episode of enhanced recent growth.

In this paper we have uncovered a direct link between the HI content
of a galaxy and its  specific star formation rate profile. This lends
further credence to the idea that galaxies where the outer disk is very blue and actively 
star-forming  compared to the inner disk have recently accreted gas.
The fact that there is no intrinsic correlation between the HI fraction and the
measured asymmetry of the optical light of the galaxy, suggests that the gas
was accreted smoothly and not in discrete units. What is
still lacking, however, is clear proof that the gas in HI-rich  galaxies has
been added to the outer disk recently. It will also   be extremely
important to obtain data that will allow us  to {\em link} the  cold gas content,
specific star formation rates  and  metallicity gradients of
disk galaxies in a self-consistent way. Only then, may we hope to fully constrain models of disk galaxy formation and chemical evolution.  

We are
currently engaged in an observational program to  obtain accurate
and homogeneous molecular gas masses for a subset of   300
galaxies from the GASS sample (Saintonge et al, in preparation;
see http://www.mpa-garching.mpg.de/COLD\_GASS/). In addition, we
are obtaining long-slit spectra for the same galaxies using the MMT
telescope (the 6.5m MMT telescope on Mt. Hopkins, AZ).  Early results indicate that specific star formation rate
and metallicity gradients in HI-rich galaxies are  indeed  tightly
correlated \citep{Moran10}.  We intend to quantify this in more detail
with larger samples of galaxies in our future work.

\section*{Acknowledgements}
We thank Y. M. Chen, L. Wang, D. Christlein, G. W. Fang, C. Li, H. Guo and H. L. Yan for help with identifying bars.
This will be the subject of a future paper. We thank L. Shao, C. Li, Y. M. Chen and S. White for useful discussions. 
XK is supported by the National Natural Science Foundation of China (NSFC, Nos. 10633020, and 10873012), the Knowledge Innovation Program of the Chinese Academy of Sciences (No. KJCX2-YW-T05), and National Basic Research Program of China (973 Program; No. 2007CB815404).

GALEX (Galaxy Evolution Explorer) is a NASA Small Explorer, launched in April 2003, developed in cooperation with the Centre National d'Etudes Spatiales of France and the Korean Ministry of Science and Technology. 

 We thank the many members of the ALFALFA team who have contributed to the acquisition and processing of the ALFALFA dataset over the last six years.  RG and MPH are supported by NSF grant AST-0607007 and by a grant from the Brinson Foundation.

Funding for the SDSS and SDSS-II has been provided by the Alfred P. Sloan Foundation, the Participating Institutions, the National Science Foundation, the U.S. Department of Energy, the National Aeronautics and Space Administration, the Japanese Monbukagakusho, the Max Planck Society, and the Higher Education Funding Council for England. The SDSS Web Site is http://www.sdss.org/.

\bibliographystyle{mn2e}

\appendix
\section{A test of the robustness of our  2-zone UV/optical colour measurements}
\label{sec:testdnuvr}
In this paper, UV/optical colours are measured by convolving the SDSS images to match the poorer resolution of the GALEX NUV images. 
In some of our galaxies, the half-light semi-minor axis $b_{50}$ becomes similar to or smaller than the size of the
$\sim 4$ pixel PSF of the GALEX NUV images. In this regime, the 2-zone colour and SFR measurements 
may no longer be valid. In addition, the GALEX MIS images are relatively shallow, so the 2-zone measurements of 
the NUV fluxes may have large errors if the signal-to-noise  is  low. 

In order to quantify these effects more precisely, we have performed simulations by transforming the images 
of nearby galaxies so that they have similar magnitudes and apparent size ranges to the galaxies
in the samples studied in this paper. 
We  select 52 galaxies from the GALEX Nearby Galaxies Survey (NGS, \citet{Gildepaz07}) with images available from the SDSS. 
We only select isolated galaxies with apparent sizes ($D_{25}$) less than 5 arcmin,  where the image is  not contaminated by foreground or
background  objects. 

We begin by measuring  $NUV-r$ and $\Delta_{o-i}(NUV-r)$ from these images (shown in purple in Figure A3).
The nearby sample turns out to have a similar distribution of these two parameters as the GASS {\em parent sample} (shown in black  on
in Figure A3). 

\begin{figure*}
 \centering
 \includegraphics[width=13cm]{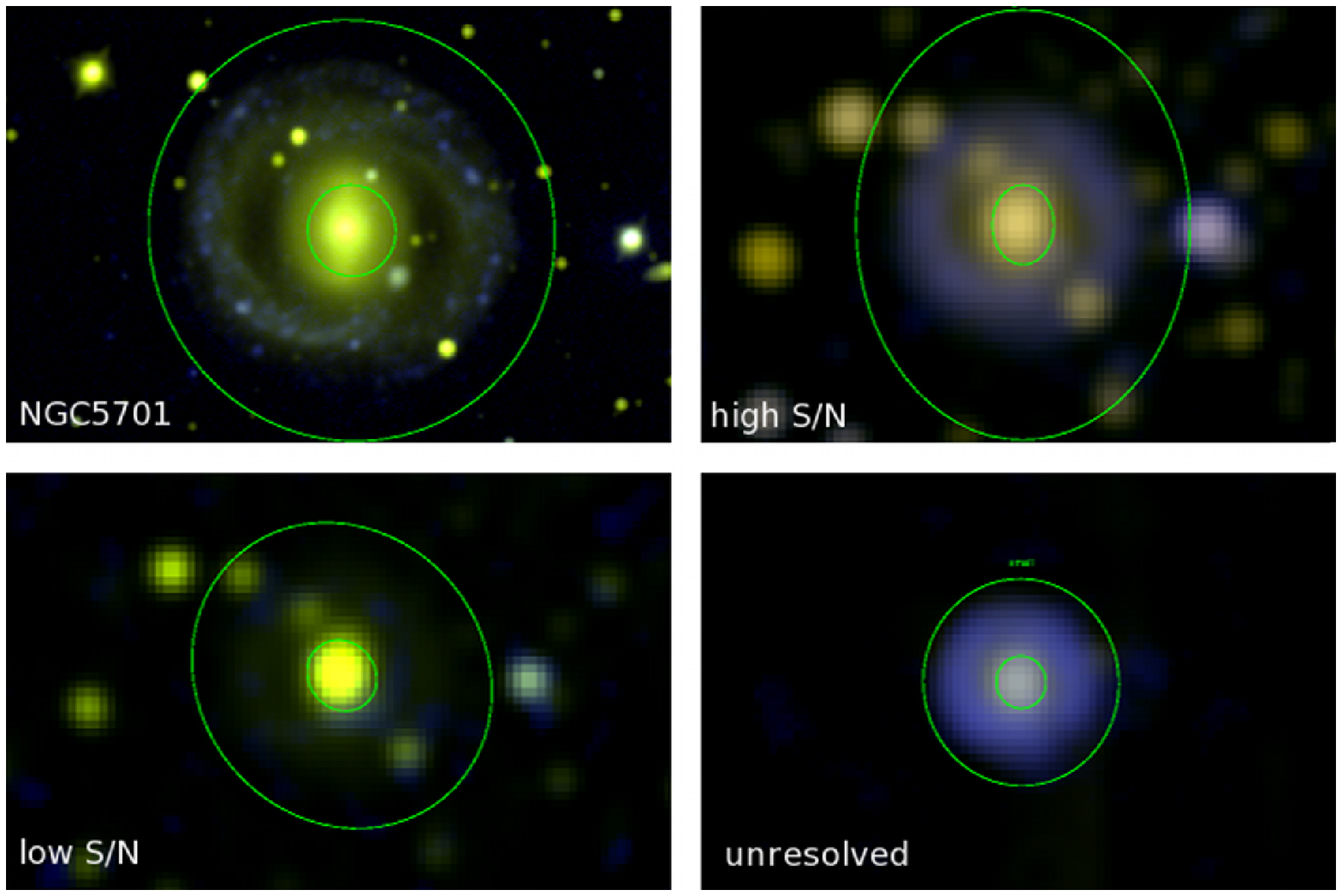}
 \caption{Original and simulated images of the galaxy NGC 5701. The left-top panel shows the original image, the top-right panel 
shows a simulated image with high $S/N$, the bottom-left panel shows a simulated image with low $S/N$, and the bottom-right 
panel shows an unresolved simulated image with high $S/N$. The two ellipses on each galaxy image mark the inner and outer regions of the galaxy.}
 \label{fig:NGC5701}
\end{figure*}

\begin{figure*}
 \centering
 \includegraphics[width=13cm]{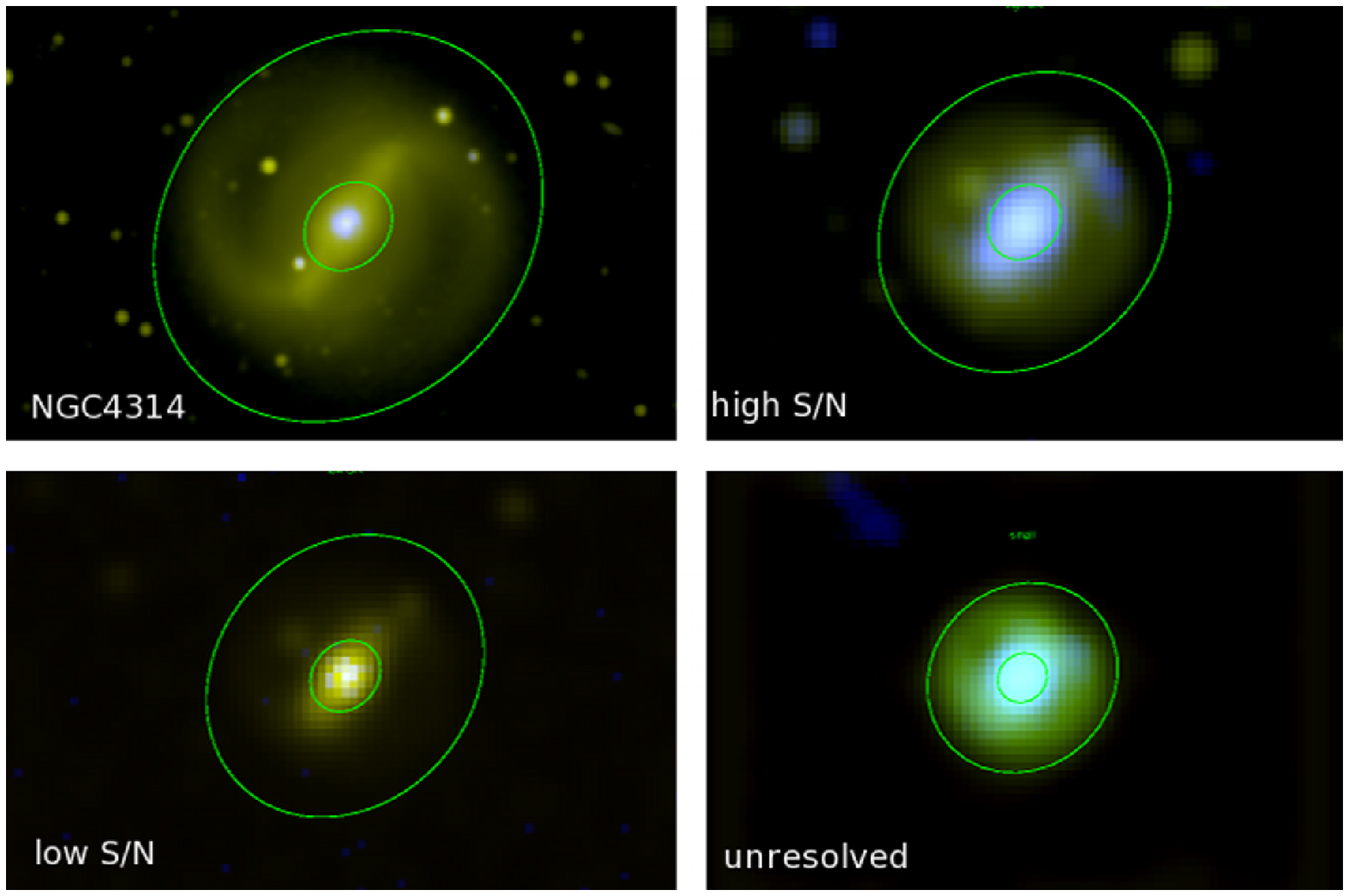}
 \caption{ As in the previous figure, except for galaxy NGC 4314, which has a blue core.}
 \label{fig:NGC4314}
\end{figure*}

We then create a library of 60,000 simulated images. For each nearby galaxy image, we select a value of $b_{50}$ and $r$-band
apparent magnitude at random from the distributions shown in black  in Figure A3.  We also
select a NUV image exposure time at random, again based on the 
real distribution of exposure times for the GASS parent sample. The NUV and $r$ band fluxes are always rescaled
so that the total $NUV-r$ colour is conserved. 
After rebinning in size, the NUV and $r$ band images are convolved with the GALEX and SDSS PSFs and
background and poisson noise are added. 
Figures~\ref{fig:NGC5701} and \ref{fig:NGC4314} show two examples of the resulting simulated images. Figure~\ref{fig:NGC5701}  
shows a typical galaxy, which is blue on the outside (with negative $\Delta_{o-i}(NUV-r)$), while 
Figure \ref{fig:NGC4314} shows a galaxy that is blue on the inside (with positive $\Delta_{o-i}(NUV-r)$).

After creating our library of 60,000 simulated images, we measure photometric parameters  
following the same steps described in Section~\ref{subsec:phot_tech}.
The distribution of  output NUV-$r$ colours, $r$-band apparent magnitudes, semi-minor axis size ($b_{50}$) and
NUV-$r$ colour gradient values are plotted as red histograms in Figure A3. We can see that the output values do
differ from the input values. 

To quantify this in more detail,     
the left panel of Figure~\ref{fig:dnuvr_criteria} shows how the difference between output and input colour gradients change as a function 
of $b_{50}$, for  galaxies with  high $S/N$ UV images ($\sigma(NUV-r)<0.1$). 
When $b_{50}>4$ pixels (which corresponds to 6 $''$), the output $\Delta_{o-i}(NUV-r)$ is
quite close to the  input value (systematically high by $\sim$0.05 mag on average). 
The scatter in the recovered gradient is less than 0.2 mag. When $b_{50}<4$ pixels, the mean value of $diff(\Delta_{o-i}(NUV-r))$ rises 
steeply and the scatter in the recovered value also increases. 
We thus impose a lower limit on  $b_{50}$ of 4 pixels when we analyze  UV/optical photometry in this paper (see Sect. 3.2).

The right panel of figure~\ref{fig:dnuvr_criteria} shows the difference between the output and input colour 
gradients  as a function of the errors on the $NUV-r$ colour, for galaxies  with
$b_{50}>4$ pixels. The average offset remains constant at $\sim$0.05 mag 
when $-$log $\sigma(NUV-r)>0.5$, and then rises steeply towards lower values of $-$log $\sigma(NUV-r)$. However, when $-$log $\sigma(NUV-r)<1$ (or $\sigma(NUV-r)>0.1$), 
the scatter in the recovered gradient  is already about 0.2 mag. This is because  the outer UV light has lower surface brightness than the inner light. 
We conclude that  $\sigma(NUV-r)<0.1$ mag should be adopted as  a lower $S/N$ limit  for robustly measuring total and 2-zone NUV-$r$ colours.

\begin{figure*}
 \centering
 \includegraphics[width=12cm]{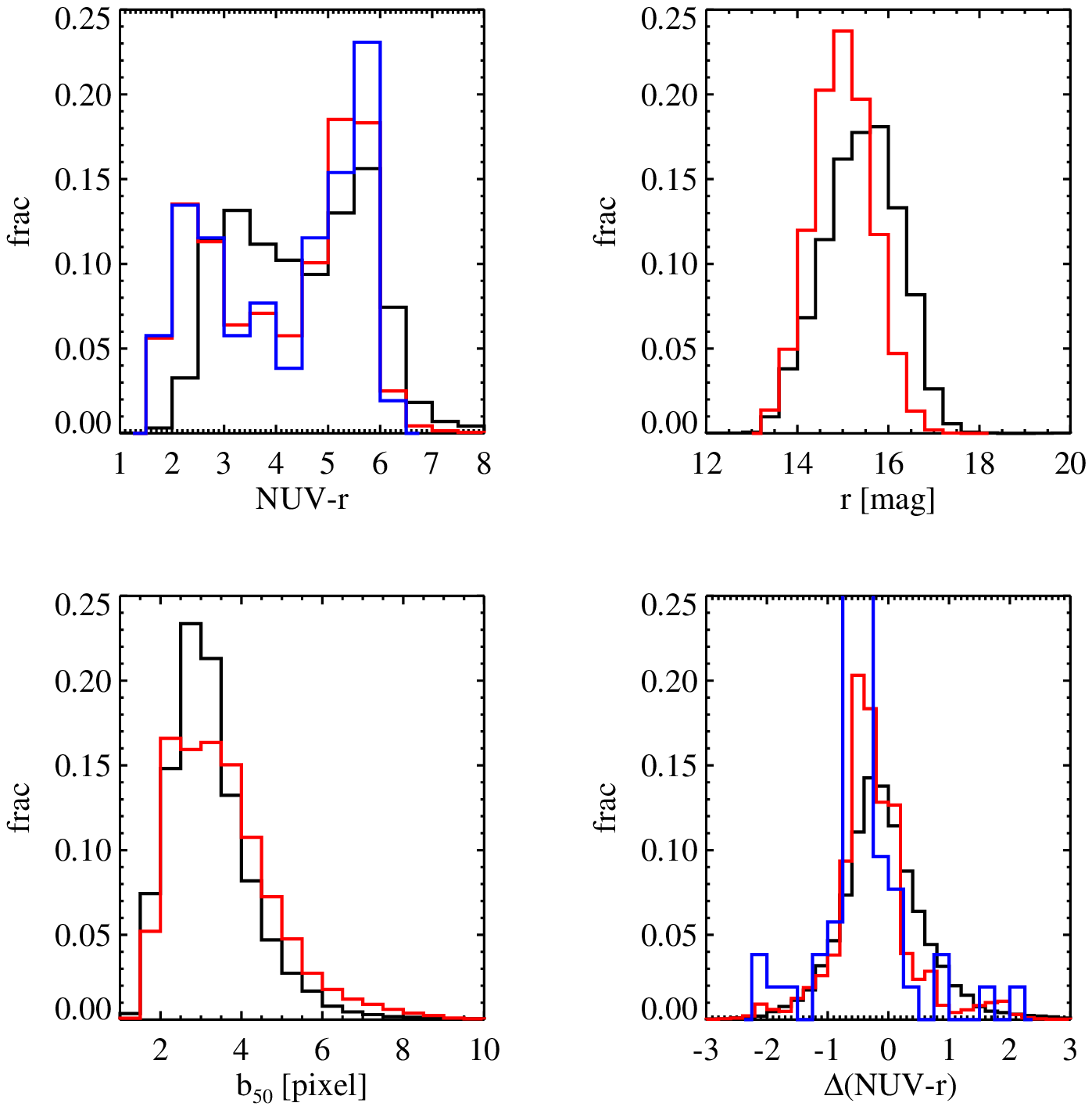}
 \caption{Distributions of  $NUV-r$ colours, $r$ band apparent magnitudes, semi-minor axis radii ($b_{50}$)
measured in the  $r$-band,  and colour differences ($\Delta_{o-i}(NUV-r)$) for the GASS parent sample (black),
for the original  $NGS$ galaxies (blue), and for the simulated images (red). }
 \label{fig:comprop}
\end{figure*}

\begin{figure*}
 \centering
 \includegraphics[width=12cm]{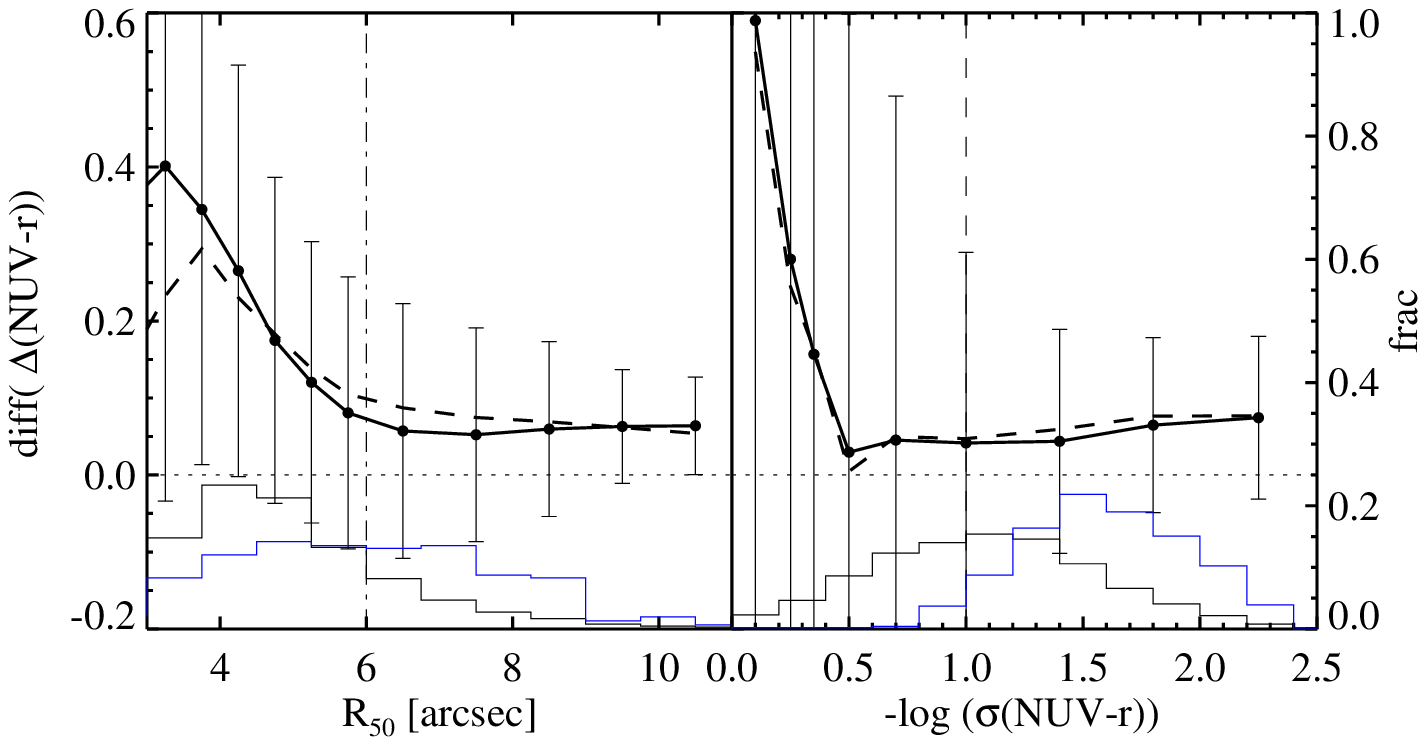}
 \caption{The average difference between the  $\Delta_{o-i}(NUV-r)$  colour for the simulated galaxy images and the original galaxy images 
is plotted as a function of half-light radius of the galaxy in arcsec (left), for cases where $\sigma(NUV-r)<0.1$.
In the right panel, the average difference is plotted  as a function of  $NUV-r$ colour errors, for cases where  $b_{50}>4$ pixels. 
The solid lines show the average difference in each $b_{50}$ bin, the dashed line shows the median, and the error bars show the r.m.s. 
 scatter in the difference around its mean value. The dotted lines mark where the offset equals 0, and the dot-dashed lines mark the location of
our recommended size and error thresholds discussed in Sect. 3.2.}
 \label{fig:dnuvr_criteria}
\end{figure*}

\end{document}